%% file: finite_energy_lc_disclinations.tex
\documentclass[12pt]{article}
\usepackage{fancyhdr}
\usepackage{isomath}
\usepackage{amsmath}
\usepackage{amsbsy}
\usepackage{amssymb}
\usepackage{amscd}
\usepackage{amsfonts}
\usepackage{verbatim}
\usepackage{euscript}
\usepackage{alltt}
\usepackage{graphicx,color}
\usepackage[font=small,format=plain,labelfont=bf,up,textfont=it,up]{caption}
\usepackage{subfigure}  
\usepackage[margin=1in]{geometry}

\input{temp-definitions}

\begin{document}
\title{A fundamental improvement to Ericksen-Leslie kinematics}
\author{ Hossein Pourmatin \  \  \ Amit Acharya\footnote{Corresponding Author: acharyaamit@cmu.edu; (412) 268 4566} \  \ \ \  Kaushik Dayal
\\ \\
Civil and Environmental Engineering \\ Carnegie Mellon University
}

\maketitle
\begin{abstract}
\noindent We demonstrate theory and computations for finite-energy line defect solutions in an improvement of Ericksen-Leslie liquid crystal theory. Planar director fields are considered in two and three space dimensions, and we demonstrate straight as well as loop disclination solutions. The possibility of static balance of forces in the presence of a disclination and in the absence of flow and body forces is discussed. The work exploits an implicit conceptual connection between the Weingarten-Volterra characterization of possible jumps in certain potential fields and the Stokes-Helmholtz resolution of vector fields. The theoretical basis of our work is compared and contrasted with the theory of Volterra disclinations in elasticity. Physical reasoning precluding a gauge-invariant structure for the model is also presented.

\end{abstract}

\section{Introduction}
While liquid crystal theory, especially that of nematics, is by now a well-established branch of condensed matter physics \cite{kleman-soft}, there is one aspect in which the classical theory may be considered deficient. Disclination line defects are an integral part of the physics of liquid crystals - however, there does not exist a classical theory, whether Oseen-Frank for statics or Ericksen-Leslie in dynamics (which naturally subsumes the static theory), that predicts bounded energy in a finite body containing a $\pm \frac{1}{2}$-strength disclination line. It is perhaps for this reason that it is commonplace amongst workers in liquid crystal theory to associate a line defect only with a singularity, e.g. the `escape' solution for a $+ 1$ strength line defect \cite{cladis-kleman} is often not referred to as a defect solution.

Director fields with infinite energy are practically problematic from the point of view of finding solutions - e.g. in statics, energy cannot be minimized in a class of functions where all competing fields with defects have undefined (infinite) energies. Practical difficulties remain in dynamics, more so if the goal is to specify dynamics of defect lines. While for the purpose of static Oseen-Frank theory involving a few defects one overcomes such problems by excluding from analysis a core region of small volume along the defect line where one assumes the theory does not hold and ascribing a finite energy to it from some unspecified-in-theory nonlinear effects, for situations where large numbers of defects may be involved, such an approach is untenable. To quote Ericksen \cite{ericksen91}
``I am interested in seeing the development of a mathematically sound theory
of defects which might be at rest, or moving. ...serious difficulties are encountered with some observed kinds of disclinations
(line defects). According to the aforementioned theories [Oseen-Frank, Ericksen-Leslie], these more violent
singularities cause energy integrals, etc., to diverge. In dealing with rather specific situations, workers have patched up the theory by excluding a tube of small
radius, assigning a finite ``core energy" to it. I just do not believe that one can use
such ideas as a basis for developing satisfactory mathematical theory, particularly
for moving defects. Such phenomena are of interest for both nematic and cholesteric
liquid crystals. In the latter, they are associated with the interesting ``blue
phases" ....."

In \cite{acharya-dayal} a theory for the dynamics of non-singular line defects has been proposed as an extension of Ericksen-Leslie \cite{leslie} theory. The model works with augmented kinematics involving the director field and an \emph{incompatible} director distortion field as a replacement for the director gradient field. The main point of departure is that the director distortion is not $curl$-free, in general. The $curl$ of the field has the kinematic meaning of an areal density of lines carrying a vectorial attribute and is referred to as the director incompatibility field. For the 1-constant Oseen-Frank energy approximation, it was shown in \cite{acharya-dayal} that the developed formalism can indeed predict the outside-core, energy density distribution of Frank's \emph{planar} director distributions for line defects with strength given by all integer multiples of $\frac{1}{2}$. Moreover, all such distributions have finite energy in the whole body (including core). The core is represented by precisely the region where the director incompatibility field is non-vanishing. However, it was not shown in that work how to extract a physically realistic continuous director field resembling Frank's planar defect fields on the whole body, and neither was it shown as to what form the director distortion tensor might take within the theory to represent planar distortion fields of disclination line defects. In this paper we achieve that goal. 

The modeling of bounded energy line defects in statics started with the pioneering work of \cite{cladis-kleman} within classical Oseen-Frank theory. The same question served as motivation for the development of the variable degree of orientation extension of classical Ericksen Leslie theory in \cite{ericksen91}. Within the De-Gennes $Q$ tensor formalism it has been pursued in \cite{sonnet-killian-hess}, \cite{kralj-virga-zumer}, \cite{Ball-Zarnescu}, \cite{majumdar_uniaxial}. In the dynamic case, representative studies are those of \cite{klein} and \cite{forest} within the $Q$-tensor approach and \cite{lin2000existence, walkington-liu} within the Ericksen-Leslie approach. In \cite{lin2000existence, walkington-liu} non-singular line-defect solutions appear due to relaxation of the unit-magnitude constraint on the director field. In contrast, in our work the demonstrated finite energy defect solutions satisfy the strict unit-vector constraint on the director field.

The defect solutions we develop here are finite energy analogs of classically accepted `energy-minimizing'  (infinite total energy) singular solutions of Oseen-Zocher-Frank theory. As is customary in the liquid crystal literature when discussing statics of defects solutions, we (almost) ignore balance of forces. What we are able to show in this regard (Section \ref{lin-mom}) is that static balance of linear momentum is satisfied outside the core(s) of disclination(s) in the absence of a body force within our model (assuming an incompressible fluid). Moreover, when there is a single disclination in the body, the natural boundary conditions for static force and moment balance representing no applied forces and couples on the boundary imply that the resultant force on any surface enveloping the disclination core vanishes. Within the core, we discuss possibilities afforded by the \cite{acharya-dayal} theory in satisfying balance of linear momentum without external body forces or flow.

We consider this work to be primarily a characterization of the kinematic augmentation of the Ericksen-Leslie theory achieved by the \cite{acharya-dayal} model, since results are not presented regarding the connection of the presented solutions to (quasi)equilibria (cf. \cite{carr1989metastable}) of the latter.

This paper is organized as follows: in Section \ref{notat} we provide notation. Section \ref{theory} describes the analytical basis of our work; Section \ref{numerics} describes the finite-element based numerical scheme employed to compute non-singular defect solutions and the results. In Section \ref{lin-mom} we discuss static balance of forces for the developed solutions, and Section \ref{conclude} ends with some concluding remarks. Two appendices provide explicit solutions that are necessary for discussing our results.
\section{Notation}\label{notat}
A ``$\cdot$" represents the inner product of two vectors, while ``$:$" represents the trace inner product of two second-order tensors, $\bfA:\bfB=A_{ij}B_{ij}$.

In rectangular Cartesian coordinates and corresponding bases and components, for a vector field $\bfA$ and a scalar field $\theta$ we write:
\begin{eqnarray}\nonumber
div\, \bfA=A_{i,i}\\\nonumber
(curl\, \bfA)_i=e_{ijk}A_{k,j}\\\nonumber
(grad\, \theta)_i=\theta_{,i}
\end{eqnarray}

For a second-order tensor field $\bfA$, we define
\[
\left(\curl\bfA\right)_{im} = e_{mjk} A_{ik,j}.
\]

The following list describes some of the mathematical symbols we use in this paper:

\begin{description}
\item[$\bfn$]: director
\item[$\bfE$]: director distortion tensor
\item[$\bfE^{\theta}$]: director distortion vector
\item[$\bfbeta$]: director incompatibility tensor
\item[$\bfbeta^{\theta}$]: director incompatibility vector
\item[$K$]: defect strength
\item[$\bfnu$]: normal vector
\item[$\bfT$]: Cauchy stress tensor
\item[$\bf\Lambda$]: couple stress tensor
\item[$\psi$]: free energy per unit mass
\item[$p$]: pressure
\item[$\theta$]: angle of the director field
\item[$\bflambda$]: layer field
\item[$\bflambda_\perp$]: incompatible part of $\bflambda$
\item[$grad\,z$]: compatible part of $\bflambda$

\end{description}
\section{Theory}\label{theory}
We consider the static (i.e. director and positional inertia-less) governing equations of the framework introduced in \cite{acharya-dayal} and demonstrate that there exists finite-energy states in the model that are solutions corresponding to line defects with planar director distributions within the Oseen-Frank constitutive assumption.

\subsection{Governing equations}\label{goveq}
Static Balance of Angular Momentum is given by the statement
\begin{equation}\nonumber
\Lambda_{ij,j} - e_{ijk}T_{jk} = 0.
\end{equation}
Ericksen's identity for the theory (i.e. a necessary condition for frame-indifference of the energy function)  is the statement
\[
e_{ijk}\left(\parderiv{\psi}{n_i} n_j + \parderiv{\psi}{E_{ir}} E_{jr} - E_{ri} \parderiv{\psi}{E_{rj}} \right) = 0.
\]
Static Balance of Linear Momentum is the statement
\[
T_{ij,j} = 0.
\]

For a free-energy density of the form $\psi\left({\bfn, \bfE}\right)$, the constitutive statements for the stress and couple stress tensors for  the theory \cite{acharya-dayal} read as
\begin{eqnarray}\nonumber
\Lambda_{ij} & = & e_{irs}n_r \parderiv{\psi}{E_{sj}}\\\nonumber
T_{ij} & = & -p\delta_{ij} - E_{ri}\parderiv{\psi}{E_{rj}},
\end{eqnarray}
where $p$ is the pressure arising from the constraint of incomressibility. Using these statements along with the Ericksen identity in angular momentum balance with some manipulations yields the fundamental governing partial differential equation of our work:
\begin{equation}\label{moment_balance}
e_{irs}\left[ n_s \parderiv{\psi}{n_r} - \left( E_{rj} - n_{r,j} \right) \parderiv{\psi}{E_{sj}} + n_r \left(\parderiv{\psi}{E_{sj}} \right)_{,j} \right] = 0.
\end{equation}
Note the interesting fact that when $E_{rj} = n_{r,j}$, i.e. there is no director incompatibility and hence no defects, we have the classical statement of angular momentum balance in statics \cite{stewart}. Inasmuch, this term in our theory may be interpreted as giving one explicit form to Ericksen's ``internal body moments'' \cite{ericksen91}, $\bfg^I$, arising from the presence of defects in the body.

We record the statement of balance of linear momentum,
\begin{equation}\label{force-balance}
-p_{,i} - \left( E_{ri} \parderiv{\psi}{E_{rj}} \right)_{, j} = 0,
\end{equation}
to be used later in Section \ref{lin-mom}.

For the sake of this paper, we consider an \emph{ansatz} consisting of director distributions parametrizable by a single angle field $\theta$ on the body and director distortion fields of the form
\begin{equation}\label{ansatz}
\bfE := \parderiv{\bfn}{\theta} \otimes \bfE^\theta,
\end{equation}
where $\bfE^\theta$ \emph{is a vector field} with possibly non-vanishing $curl$. When there are no line defects, we require $\bfE^\theta = grad \,\theta$ so that $\bfE = grad\,\bfn$.

Let the \emph{core} region for a given $\bfE^\theta$ field on the body be the set of points on which 
\[
\bfbeta^\theta := curl \,\bfE^\theta \neq \bf0
\]
and 
\[
\bfbeta := curl (\bfE - grad\,\bfn) \neq \bf0.
\]
We refer to an $\bfE^\theta$ field as containing an isolated line defect if its $\bfbeta^\theta$ field is non zero only in a cylindrical region that forms a closed loop or extends from one boundary of the body to another.

 The goal now is to construct pairs of $\left( \bfE, \bfn \right)$ fields
\begin{enumerate}
\item that satisfy (\ref{moment_balance});
\item that produce Frank's isolated line defect-like energy density fields outside the core region, while producing bounded energy in finite bodies (including the core region), when no moments are applied on the boundary of the body;
\item that produce Frank's wedge disclination-like director distributions except the core and possibly another region of `small', but non-zero, volume in the body;
\item for which the line integral of $\bfE^\theta$, along any closed contour surrounding the core of an isolated line defect of strength $K$, evaluates to $2 \pi K$. 
\end{enumerate} 
\subsection{Construction of bounded energy distortion and director solutions}\label{argument}
We work with a rectangular Cartesian coordinate system with unit vectors $\bfe_i, i = 1,2,3$. $\theta$ represents the angle of the director measured counter-clockwise from the $x_1$-axis in the $x_1-x_2$ plane (looking down the $x_3$-axis). Thus
\begin{equation}\label{n}
\bfn = \cos\theta \bfe_1 + \sin\theta \bfe_2.
\end{equation}
For the sake of illustrating the essentials of our approach, we work with the 1-constant Oseen-Frank energy density approximation
\begin{equation}\label{1-const}
\psi\left(\bfn, \bfE \right) = \frac{1}{2} \kappa \bfE : \bfE = \frac{1}{2} \kappa E_{ij}E_{ij} = \frac{1}{2} \kappa E^\theta_i E^\theta_i
\end{equation}
and
\begin{equation}\label{1-const-mom}
\parderiv{\psi}{E_{sj}} = \kappa E_{sj} = \kappa \parderiv{n_s}{\theta} E^{\theta}_{j}.
\end{equation}
Noting
\[
E_{rj} - n_{r, j} = \parderiv{n_{r}}{\theta} \left(E^\theta_j - \theta_{,j} \right)
\]
and (\ref{1-const})-(\ref{1-const-mom}), (\ref{moment_balance}) reduces to
\[
e_{irs} n_r \kappa \left[ \parderiv{n_s}{\theta} E^\theta_j \right]_{,j} = 0.
\]
Considering (\ref{n}) and realizing that $e_{irs} n_r \parderiv{^2 n_s}{\theta^2} \theta_{,j} = 0$, static angular momentum balance reduces to
\[
e_{irs} n_r \parderiv{n_s}{\theta}E^{\theta}_{j,j} = 0
\]
which is satisfied if and only if
\begin{equation}\label{divEtheta}
div \bfE^{\theta} = 0.
\end{equation}

To fulfill the fourth requirement on the field $\bfE^{\theta}$, let us now consider a distribution of $\bfbeta^\theta := curl \bfE^\theta$ in a right-cylinder parallel to $x_3$ such that
\begin{equation}\label{beta}
\int_{A} \bfbeta^\theta \cdot \bfe_3\  dx_1dx_2 = 2\pi K,
\end{equation}
where $A$ is the set of points representing the cross section of the core cylinder. Thus, in addition to (\ref{1-const-mom}), we require
\begin{equation}\label{curlEtheta}
curl \bfE^\theta = \bfbeta^\theta.
\end{equation}
Finally, to impose the fact that we seek non-trivial solutions for an unloaded body, we impose vanishing moments on the boundary of the body which, under the current ansatz is satisfied if
\begin{equation}\label{bc}
\bfE^\theta \cdot \bfnu = 0 
\end{equation}
on the boundary of the body with normal field $\bfnu$.

It is straightforward to see that, for a specified field $\bfbeta^\theta$ on a simply-connected domain, (\ref{divEtheta})-(\ref{curlEtheta})-(\ref{bc}) has a unique solution. In \cite{acharya-dayal}, these equations are solved explicitly for a radially symmetric $\bfbeta^\theta$ distribution (see Appendix \ref{appB}), and the outside-core result shown to be exactly the same as the gradient of the Frank angle field for straight wedge disclinations. In the general case, (i.e. not necessarily radially-symmetric $\bfbeta^\theta$ field, but still localized in core cylinder distributions) we would now like to extract the outside-core result into a gradient of a \emph{continuous everywhere} scalar angle field $\theta$ which will serve to define the $\bfn$ field of the pair $\left( \bfE, \bfn \right)$ describing an isolated defect. Of course, in the process we would like to have $\bfE^\theta$ still be the solution to (\ref{divEtheta})-(\ref{curlEtheta})-(\ref{bc}) and $\bfE^\theta = grad\,\theta$ when $\bfbeta^{\theta} = \bf0$.

To perform this extraction, suppose for a moment it is possible to construct a square-integrable vector field $\bflambda$ such that
\begin{equation}\label{curllambda}
curl\,\bflambda = - \bfbeta^\theta
\end{equation}
on the body, where the $\bfbeta^\theta$ field is identical to the one occurring in (\ref{curlEtheta}). Then a Stokes-Helmholtz resolution of the field $\bflambda$ can be written as
\[
\bflambda = \bflambda_\perp + grad\,z
\]
with
\begin{equation}\label{lambda-beta}
curl\, \bflambda_\perp = curl\, \bflambda = - \bfbeta^\theta
\end{equation}
\[
div\, \bflambda_\perp = 0
\]
and
\[
\bflambda_\perp \cdot \bfnu = 0 \ \  \text{on boundary of body,}
\]
and because of the uniqueness of solutions to (\ref{divEtheta})-(\ref{curlEtheta})-(\ref{bc}), $\bfE^\theta = - \bflambda_\perp$.

\begin{figure}
\begin{center}
\includegraphics[height=3in,width=4in,angle=0]{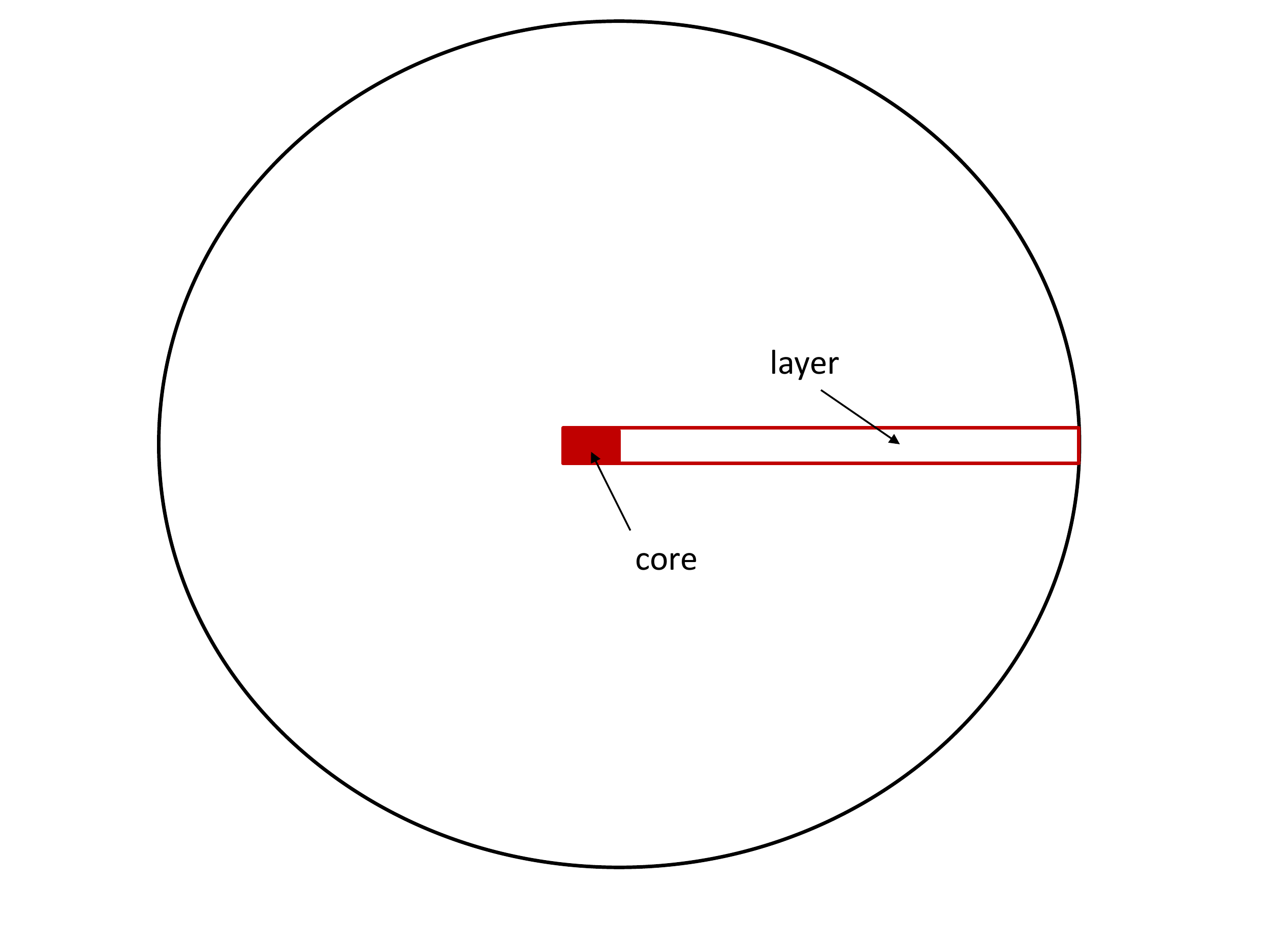}
\caption{\label{layer_core} Schematic of cross-section of a 3-d body showing layer and core geometry.}
\end{center}
\end{figure}

Suppose further now that the field $\bflambda$  \emph{is non-vanishing only on a layer-like region of non-zero volume}\footnote{ Cf. \cite{Dewit}; DeWit considers a distributional layer field in the context of disclinations in solids related to defects in positional order.} that may be visualized as a terminating fattened 2-d surface which, moreover, contains as a proper subset the core cylinder on which $\bfbeta^\theta$ has support (Figure \ref{layer_core}). 

Then, as an immediate consequence we have that
\[
 - grad\,z = \bflambda_\perp = - \bfE^\theta \ \ \ \text{outside the layer}.
\]
Thus, if we were to now declare the potential $z$ as the required angle-field $\theta$, \emph{we would have completed the needed extraction}.

In practice (Section \ref{numerics}), we implement the above idea by defining
\begin{equation}\label{Etheta}
\bfE^\theta := grad\,\theta - \bflambda
\end{equation}
and requiring this combination to satisfy (\ref{divEtheta})-(\ref{curlEtheta})-(\ref{bc}) so that
\begin{equation}\label{gradtheta}
grad\, \theta - \bflambda = - \bflambda_\perp
\end{equation}
is enforced by uniqueness implying $grad\, \theta = grad\,z$. When accompanied by the requirement that $\bflambda = \bf{0}$ if $\bfbeta^\theta = \bf{0}$,  the strategy ensures that $\bfE = grad\,\bfn$ in the absence of defects in the body. We note the important fact that the Stokes-Helmholtz decomposition of an $L_2$ vector field implies that $grad\,z$ is square-integrable and therefore $grad\,\theta$ cannot have a non-square-integrable singularity in the body. \emph{The set of Definitions (\ref{ansatz}) and (\ref{Etheta}) constitutes a primary result of this paper}.

So, the only question that remains is whether such a layer-field $\bflambda$ satisfying (\ref{curllambda}) can in fact be constructed. As we show in Section \ref{numerics}, it suffices to consider the layer region as in Figure \ref{layer_core}, prescribe an appropriate constant vector field (representative of an interpolation of the required jump in the value of $\theta$, i.e. $2 \pi K$, in the direction transverse to the  layer) for the values of $\bflambda$ outside the core in it, and tapering this interpolated jump to zero over the width of the core. Outside the layer, $\bflambda$ is assigned to vanish. Since $\bflambda$ represents a transverse `gradient' in the layer whereas in the core this transverse variation has a gradient in an in-plane direction, it cannot have vanishing $curl$ in the core. Thus we represent a non-singular, but localized, core of a line defect. 

We now ask the question of the region where $\bfbeta$ is non-vanishing corresponding to an isolated wedge disclination. The existence of a differentiable extracted $\theta$ field (which can be arranged as can be seen from the construction in Appendix \ref{appA}) implies that the $\bfbeta$ field corresponding to an isolated defect would be non-vanishing at most in the layer, since $grad\,\theta = \bfE^\theta$ outside the layer and
\begin{equation}\label{beta-in-layer}
E_{rj,k} = \parderiv{^2 n_r}{\theta^2} \theta_{,k} E^{\theta}_{j} + \parderiv{n_r}{\theta} E^\theta_{j,k}.
\end{equation}

Our use of a layer field may be interpreted as a regularized analog of the possible terminating discontinuity in the displacement field over a `cut' surface in the Weingarten-Volterra (WV) process of elastic dislocation theory, adapted to a much simpler situation than elasticity; thus, the construction shows direct links between the Stokes-Helmholtz (SH) resolution of a vector field (in this instance, the layer field) and the Weingarten-Volterra process. In particular, how the `$curl$' part of SH encodes, through a smooth field, topological content that can only be represented by a nasty singularity in the gradient of a discontinuous potential field in the WV process. And, how the `gradient' part of SH of an appropriately designed vector field (the layer field) can represent most of the characteristics of the discontinuous potential field of the WV process away from the discontinuity.

\subsection{Does topological defect density determine the director distribution?}
The construction above lays bare an interesting fact. Note that $\bfE^\theta$ is uniquely determined by the defect density field $\bfbeta^{\theta}$ (and balance of moments) but $grad\,\theta$, and hence the predicted director distribution, is defined by the compatible part $grad\,z$ of $\bflambda$. An alternative way of seeing this is to substitute
\[
\bfE^{\theta} =  grad\,\theta - \bflambda_\perp - grad\,z
\]
into (\ref{divEtheta})-(\ref{curlEtheta})-(\ref{bc}) (this idea is not restricted to the 1-constant energy in any way). Thus, while the energy density distribution, and therefore the couple stress, can be correctly predicted purely from the knowledge of the defect distribution (in this static setting), uniquely predicting the director distribution requires additional physical input in the presence of defects. In particular, the layer-like $\bflambda$ field consistent with a specified isolated defect density field orients the nonsingular director distribution of the defect, as is shown in Section \ref{numerics}. To draw an association with classical potential theory, specifying the surface of discontinuity of a scalar potential field whose gradient is required to match, except on the discontinuity surface, a prescribed, smooth, irrotational vector field on a doubly-connected region containing a toroidal or through-hole eliminates the vast non-uniqueness associated with such definition of a potential otherwise. The argument also reveals why the classical theory with only the director field as a degree of freedom is necessarily limited  in the context of modeling defects.

It is also interesting to observe that this situation is entirely analogous to the extension that was required for making a prediction of permanent deformation \cite{Acharya2001} in the elastic theory of continuously distributed dislocations \cite{kroner}, \cite{willis}. Comparing and contrasting with prevalent notions in gauge theories, $\theta$ and the total displacement field of plasticity theory are not inconsequential fields to be gauge-transformed away \emph{even} when they have no energetic cost through the Lagrangian in the presence of gauge fields ($\bflambda$, in this context), but direct physical observables that participate in understanding energy dissipation, shape changes, and optical response of the body. Hence, their evolution requires additional physical specification beyond the topological defect density field. 
\subsection{Relationship with disclinations in elasticity theory}\label{volterra}
Our construction of the $\bflambda$ field draws direct motivation from a standard question of potential theory related to understanding the allowable jumps of a scalar potential across a 2-d surface in a  multiply-connected domain described by a through-hole in a 3-d body. For example, consider Figure \ref{layer_core} with the layer shrunk to zero thickness and also the consequent core to a line of puncture through the plane of the paper; the core-line represents the through-hole. The 2-d surface is such that a cut along it renders simply-connected the body with the hole. Additionally, the gradient of the potential field is required to match a prescribed irrotational vector field on the simply-connected domain obtained through the cut. 

We note here that the question above is much simpler in its details than the question posed by Weingarten and Volterra related to understanding the jumps of a displacement (vector) field, when its symmetrized gradient is required to match a prescribed \emph{symmetric} tensor field which satisfies the St.-Venant compatibility condition of linear elasticity theory in the simply-connected domain resulting from a planar cut of the multiply-connected domain with the hole. Indeed, in this latter case, a key construction is a skew-symmetric tensor field (infinitesimal rotation) from the prescribed symmetric tensor field, whose jump across the cut-surfaces (that serve as parts of the boundary of the simply-connected domain) relates directly to the disclinations of linear elasticity theory. It should be clear, then, that in the liquid crystal theory context, even if one chose to discuss the issue of allowable jumps in the director field context rather than the angle parametrization field, there is no natural way a relevant symmetric tensor field arises from the director gradient whose unique\footnote{Up to a spatially uniform skew-symmetric tensor field.} infinitesimal rotation field in a simply-connected domain may then be used to define the notion of an isolated singular disclination in liquid crystal theory. Given that the Oseen-Frank and Ericksen-Leslie theories are geometrically exact, the situation is not very different if a link is attempted between liquid crystal disclinations and the singular defects that characterize the  jumps in finite rotation associated with an elastic strain field that satisfies the finite strain compatibility conditions in a multiply-connected domain. Again, the issue crucially hinges on the question being posed in terms of the basic tensor field being a positive-definite symmetric second order tensor field (the Right-Cauchy-Green tensor field) and there does not seem to be any natural way in which the director gradient may be considered to generate such a field. Such an observation is consistent with that made in \cite{kleman-friedel} (Sec. I.E.1, p. 67): ``This explains the interest in reconsidering the continuous theory of defects, although new concepts have to emerge. The case of mesomorphic phases requires an extension of the theory of continuous defects for solids to situations where there is locally only one physical direction (the director), i.e. no local trihedron of directions, as in the uniaxial nematic $N$, the Sm$A$, and the columnar $D$ cases." We believe that our work provides the natural extension of the theory of continuous defects in solids to the mesomorphic phases named above.

As a further example of such differences, we note, as mentioned in \cite{kleman-friedel} Sec. IIA (p. 69, Remark), that the Volterra process is properly defined only for magnitude of rotation angles less than $\pi$. However, as far as wedge disclinations in liquid crystals are concerned the $+\frac{1}{2}$ and $+\frac{3}{2}$ strength defects are entirely different entities - as shown in Section \ref{numerics}, our theory is able to predict such differences.
\section{Computation of bounded energy distortion and director solutions}\label{numerics}

We demonstrate and evaluate the ideas presented in Section \ref{argument} through explicit constructions for different kinds of defects. First, we present results for straight wedge and twist disclinations and compare our results with Frank's planar model \cite{Frank}. Then we take the next step forward and simulate twist disinclination loops.

\subsection{Straight disclinations}\label{wedge}
We refer to disclinations with core cylinder along a straight line, as opposed to a general curve or a closed loop, as {\em straight disclinations}. A straight wedge disclination is one for which the defect line is along the axis of rotation of the director field. In the case of a twist disclination, the line is perpendicular to the axis of rotation.

\subsubsection{Wedge disclinations}
Although it is uncommon find an isolated wedge disclination in nature due to its considerable energy cost, it is nevertheless a canonical problem for theoretical tools and we study such a 2-dimensional line defect \cite{kleman-soft}.

Consider $\theta$ as a function of the coordinates $x_1$ and $x_2$, so that the director field would be planar. According to (\ref{beta}), (\ref{curllambda}) and Stokes' theorem
	\begin{align}\label{criterion}
	 \int\beta_3^{\theta}dA=- \int\curl\bflambda\cdot\bfe_3 dA=- \oint\bflambda\cdot dr=2\pi K.
	 \end{align}
This criterion indicates that the line integral of the vector field $\bflambda$ over any closed curve bounding the core is non-vanishing. Inspired by the Weingarten-Volterra process in elastic dislocation theory and adapting it to the simpler situation at hand as indicated in Sections \ref{argument}-\ref{volterra}, we define $\bflambda$ as a vector field with support on a thin layer that originates from the defect core and extends all the way to the boundary:
\begin{equation}\label{sing_wall}
\bflambda:= - \frac{2\pi K}{2r_1}\bfe_1\ \ \ \ \text{for  }\left\{
\begin{array}{ll}
x_2>0\\
|x_1|<r_1.
\end{array}\right.
\end{equation}
Here, $r_1$ defines the spatial width of the layer. The derivative of $\bflambda$ with respect to $x_2$ is zero everywhere, except at the layer boundary at $x_2 = 0$. Consequently, $\bfbeta^{\theta}$ will be a singular wall-like distribution of finite extent supported on this layer boundary. However, for reasons indicated in Section \ref{argument}, the total energy of the disclination so defined would still necessarily be finite.

In agreement with Figure \ref{layer_core}, we now endow this singular wall with a finite width\footnote{Of course, ideally one would seek such a width as an outcome of the full dynamical model. As a first step, we show here that isolated disclinations of specified strength with finite-size cores can be solutions to balance of moments in our model.}.
Thus, the following definition for the layer field is used:
\begin{equation}\label{lambda-def}
\lambda_1:=\left\{
\begin{array}{ll}
\frac{2\pi K}{2r_1} & x_2>\frac{r_0}2\\
\frac{2\pi K}{2r_1}\frac{r_0+2x_2}{2r_0} & |x_2|\le\frac{r_0}2 
\end{array}\right.
|x_1|<r_1.
\end{equation}
Figure \ref{lambda-beta} shows the norm of $\bflambda$ and the resulting $\bfbeta^\theta$ field from it, where both fields have been normalized by $K$:

\[
\tilde{\bflambda} = \frac1K\bflambda \ \ \ \ \ \tilde{\bfbeta^{\theta}} = \frac1K\bfbeta^{\theta}.
\]

\begin{figure}[!htb]
\centering{
	\subfigure[$|\tilde{\bflambda}|$] {
		\includegraphics[width=65mm,trim=2.cm 2.8cm 2.7cm 2.8cm, clip=true]{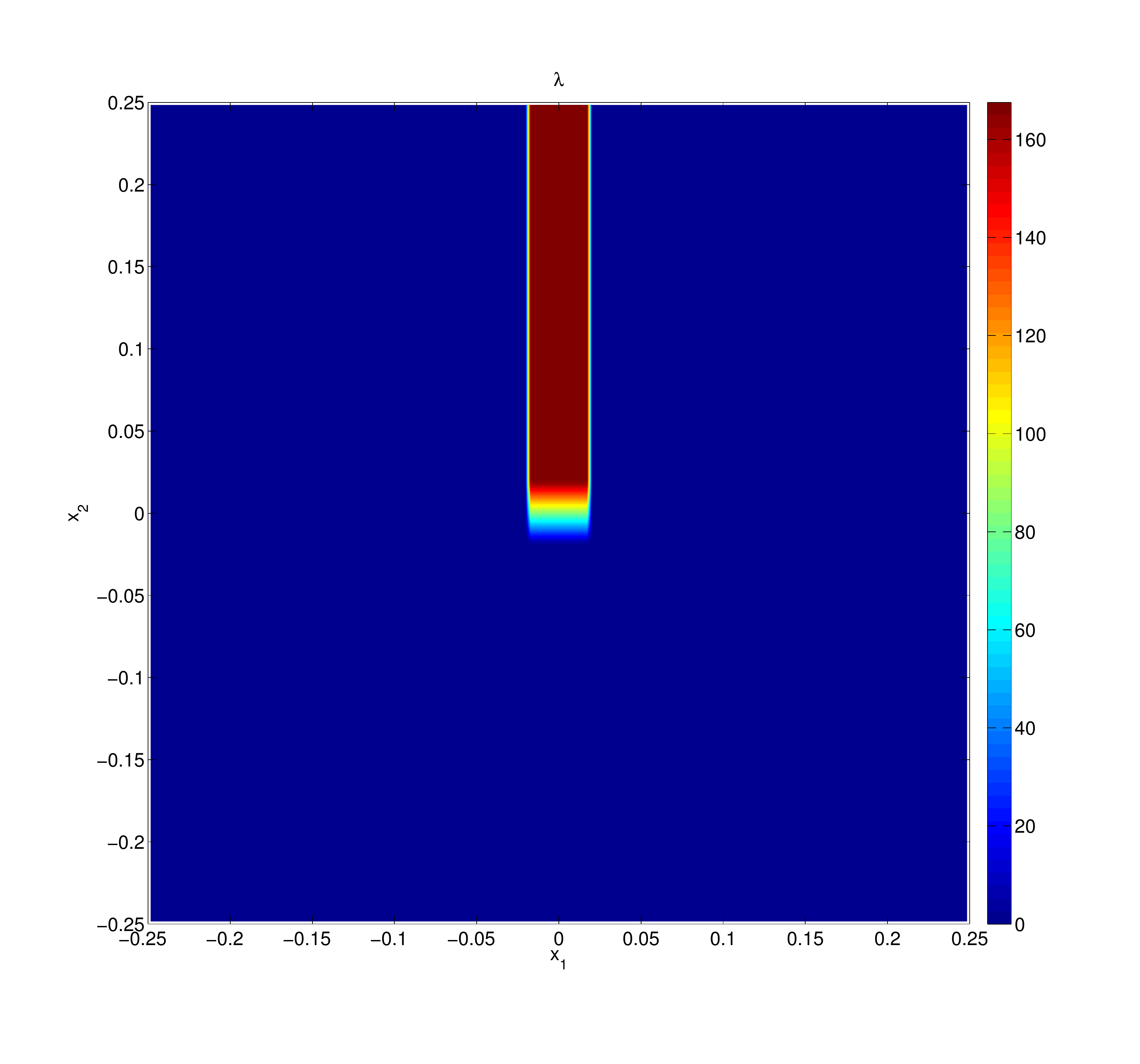}
	}
	\subfigure[$|\tilde{\bfbeta^{\theta}}|$] {
		\includegraphics[width=65mm,trim=2.cm 2.8cm 2.7cm 2.8cm, clip=true]{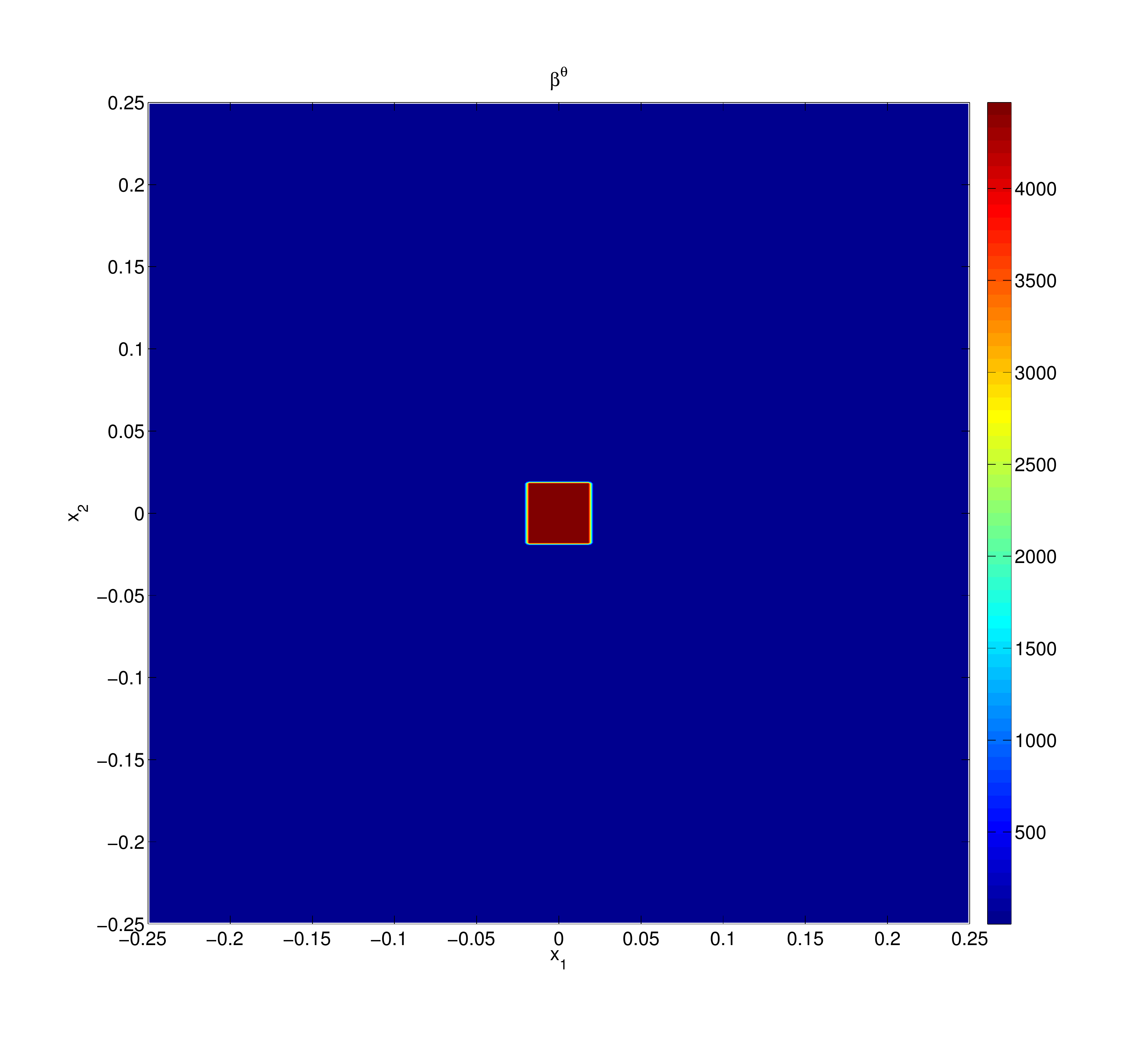}
	}
}
\caption{Normalized $\bflambda$ field, resulting from (\ref{lambda-def}) and the corresponding $\bfbeta^{\theta}$ field.}
\label{lambda-beta}
\end{figure}

\begin{figure}[!htb]
\centering{
	\subfigure[$|\tilde{\bfbeta}|$] {
		\includegraphics[width=65mm,trim=2cm 2.8cm 2.4cm 2.8cm, clip=true]{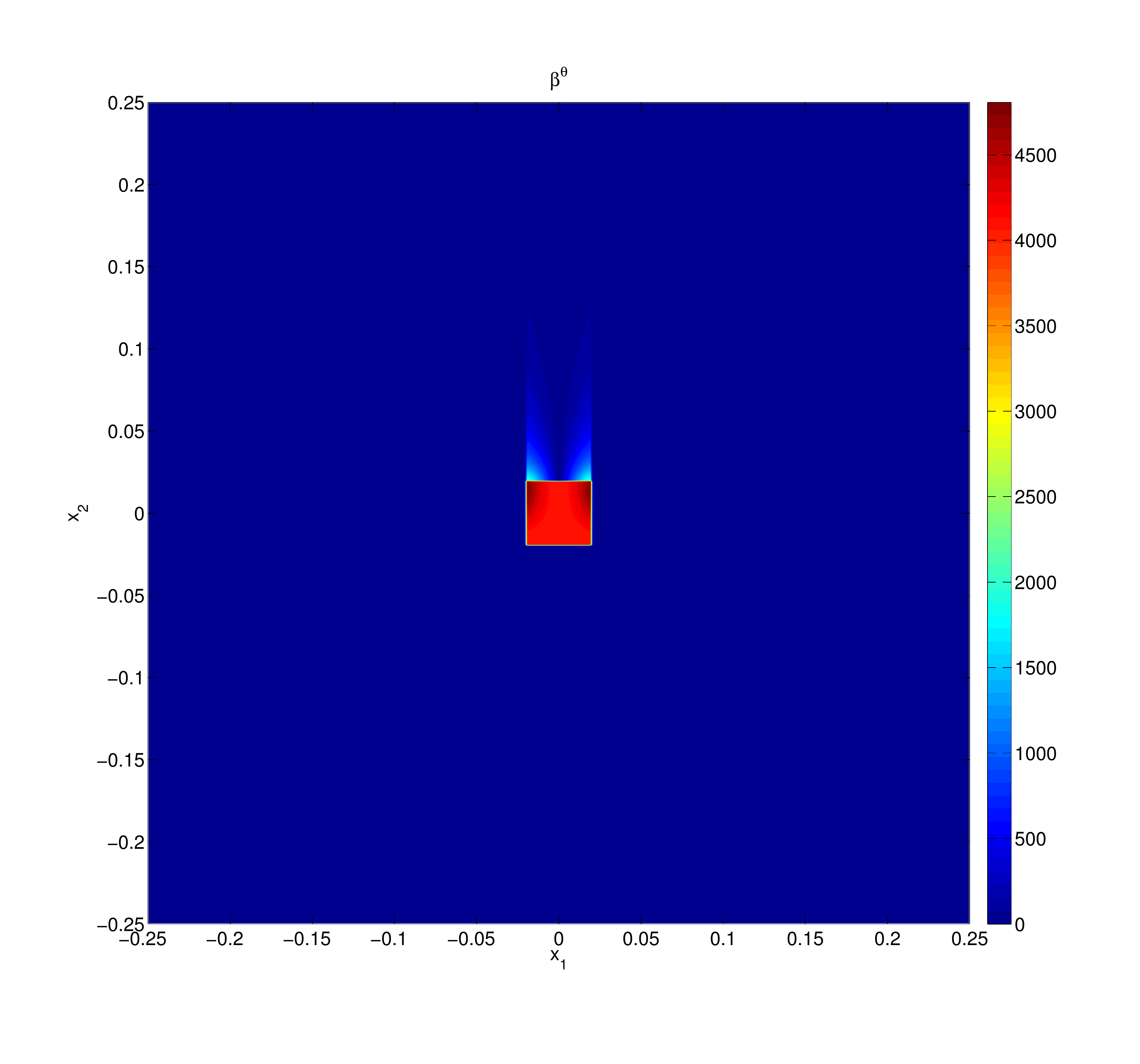}
	}
	\subfigure[$\log(|\tilde{\bfbeta}|)$] {
		\includegraphics[width=65mm,trim=2cm 2.8cm 2.7cm 2.8cm, clip=true]{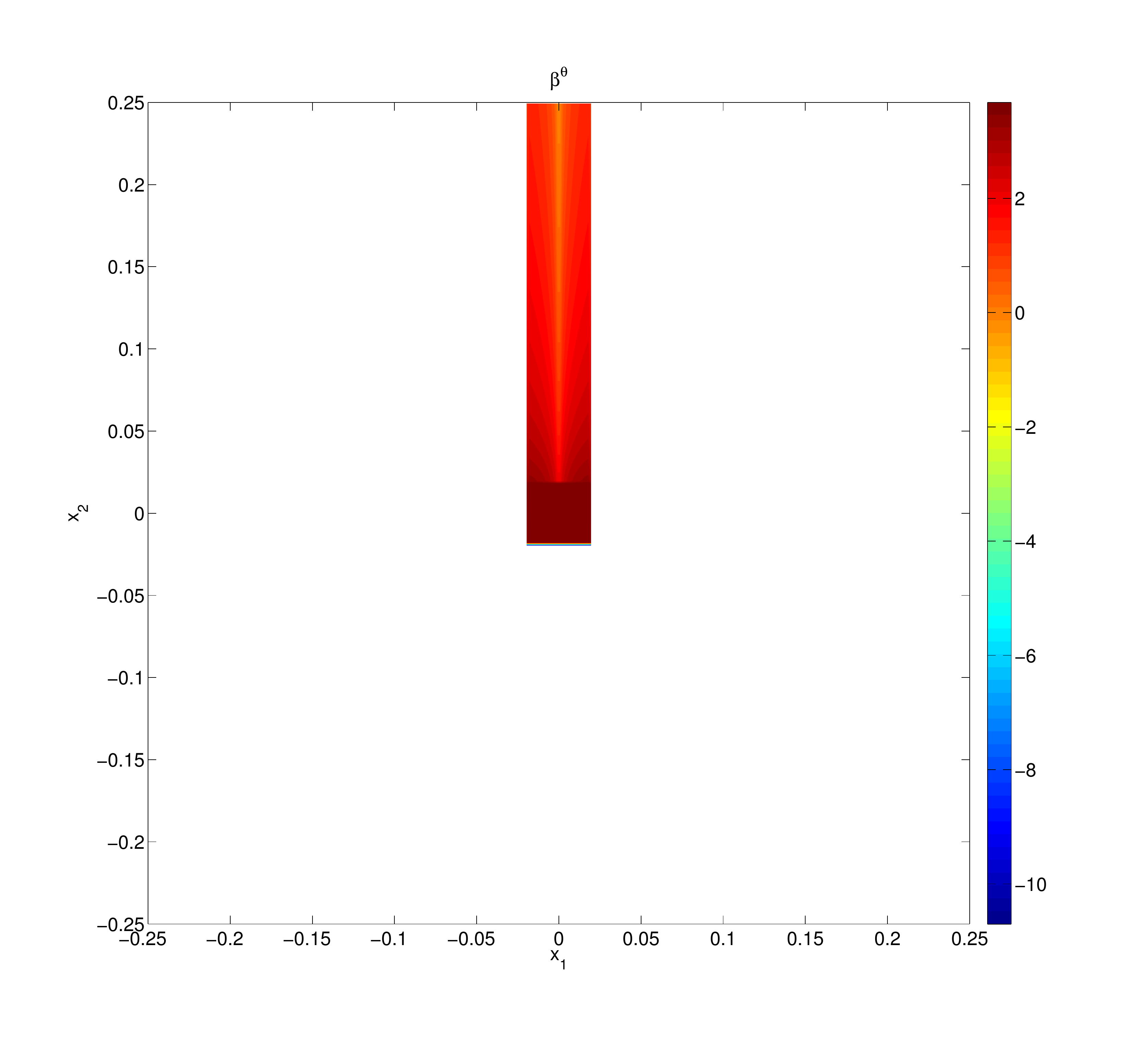}
	}
}
\caption{Normalized $\bfbeta$. The affects of the first term in (\ref{beta-term}) are more obvious in the logarithmic plot. The white area in part (b) represents zero for $\tilde{\bfbeta}$.}
\label{norm-beta}
\end{figure}

Note that, as mentioned in section \ref{argument}, the $\bfbeta$ field may not be as localized as $\bfbeta^{\theta}$ since
\begin{align}\nonumber
\bfbeta&=\curl(\bfE)\\ \nonumber
\beta_{ij}&=e_{jkl}E_{il,k}\\ \nonumber
&=e_{jkl}\big(\frac{\partial n_i}{\partial\theta}E_l^{\theta}\big)_{,k}\\ \nonumber
&=e_{jkl}\big(\frac{\partial^2 n_i}{\partial\theta^2}\theta_{,k}E_l^{\theta}+\frac{\partial n_i}{\partial\theta}E_{l,k}^{\theta}\big)\\ \nonumber
&=e_{jkl}\big(\frac{\partial^2 n_i}{\partial\theta^2}\theta_{,k}(\theta_{,l}-\lambda_l)+\frac{\partial n_i}{\partial\theta}E_{l,k}^{\theta}\big),
\end{align}
and noting that $e_{jkl}E_{l,k}^{\theta}=\beta_{j}^{\theta}$ and $e_{jkl}\theta_k\theta_l=0$,

\begin{align}\label{beta-term}
\beta_{ij}&=-e_{jkl}\big(\frac{\partial^2 n_i}{\partial\theta^2}\theta_{,k}\lambda_l\big)+\frac{\partial n_i}{\partial\theta}\beta_j^{\theta}.
\end{align}
One can easily see that while the second term is localized inside the defect core, the first term is not necessarily zero inside the layer. However, since $\theta_{,k}$ decays sharply from the defect core, the effect of this term vanishes rapidly away from the core. Figure \ref{norm-beta} shows this deviation from being a perfectly localized field.

Equations (\ref{divEtheta}), (\ref{gradtheta}), and (\ref{lambda-def}) yield the director field from the following Poisson's equation
\begin{equation}\label{Poisson}
	div\left(grad\,\theta\right)=div\,\bflambda.
\end{equation}

As a close reference point for the numerically computed approximate director solutions that are generated subsequently, the closed-form solution of  (\ref{Poisson}) in an infinite domain is derived in the Appendix \ref{appA} corresponding to smooth approximating sequences to the discontinuous layer fields, i.e. (\ref{sing_wall})-(\ref{lambda-def}), we consider. We have used the Galerkin finite element method to solve (\ref{Poisson}).

In order to compare our energy and director field results with those available in the literature (e.g. \cite{kleman-soft}), we will consider {\em line wedge disclinations in an infinite space}. To compare the results with Frank's planar model, we consider the following equation with Neumann boundary condition:
\[
\left\{\begin{array}{ll}
div\,grad\,\theta = div\,\bflambda\ \ \ \ \ \text{on }V\\
grad\,\theta\cdot\bfnu = grad\,\phi\cdot\bfnu \ \ \ \text{on }\Gamma,
\end{array}\right.
\]
where $\Gamma$ is the exterior boundary of the finite domain $V$ and $\phi$ is the angle of the director field with the $x_1$ axis in Frank's solution \cite{Frank}: 

\begin{equation}\label{frank}
\phi = K \tan^{-1}\Big(\frac{x_2}{x_1}\Big)+c
\end{equation}
with $c$ a constant. For the purpose of evaluating the boundary condition for the domains involved, it suffices to consider $grad\,\phi$ to be given by
\begin{equation}\label{frank-grad}
\frac K{r^2}\left( -x_2 \bfe_1 + x_1 \bfe_2\right)
\end{equation}
(to the neglect of a surface Dirac distribution related to the discontinuity in Frank's solution (\ref{frank})). Moreover, one must assign the value of $\theta$ at one point in the domain for uniqueness of solutions. Note that this assignment is tantamount to fixing $c$ in Frank's solution. We choose to assign a value $\theta_0$ at one point on the right-hand-side boundary of the layer/strip on which $\bflambda$ has support. 

Now, using this Neumann boundary condition, the weak form of (\ref{Poisson}) for use in the Galerkin finite element method is

\begin{equation}\label{weak-form}
\int_V\delta\theta_{,i}\left(\theta_{,i}-\lambda_i\right)dV-\int_{\Gamma}\delta\theta\phi_{,i}\nu_idS=0.
\end{equation}
Here $\delta\theta$ is a continuous test function for $\theta$. Bilinear, 4-node, quadrilateral elements are used for our simulations. Needless to say, the resulting $\theta$ will be a continuous field and we show converged results for the director field with mesh refinement in Figure \ref{error-director}. 

\begin{figure}[!htb]
\centerline{
	\subfigure[$K=\frac12$] {
		\includegraphics[width=50mm,trim=3.6cm 2.8cm 3cm 2cm, clip=true]{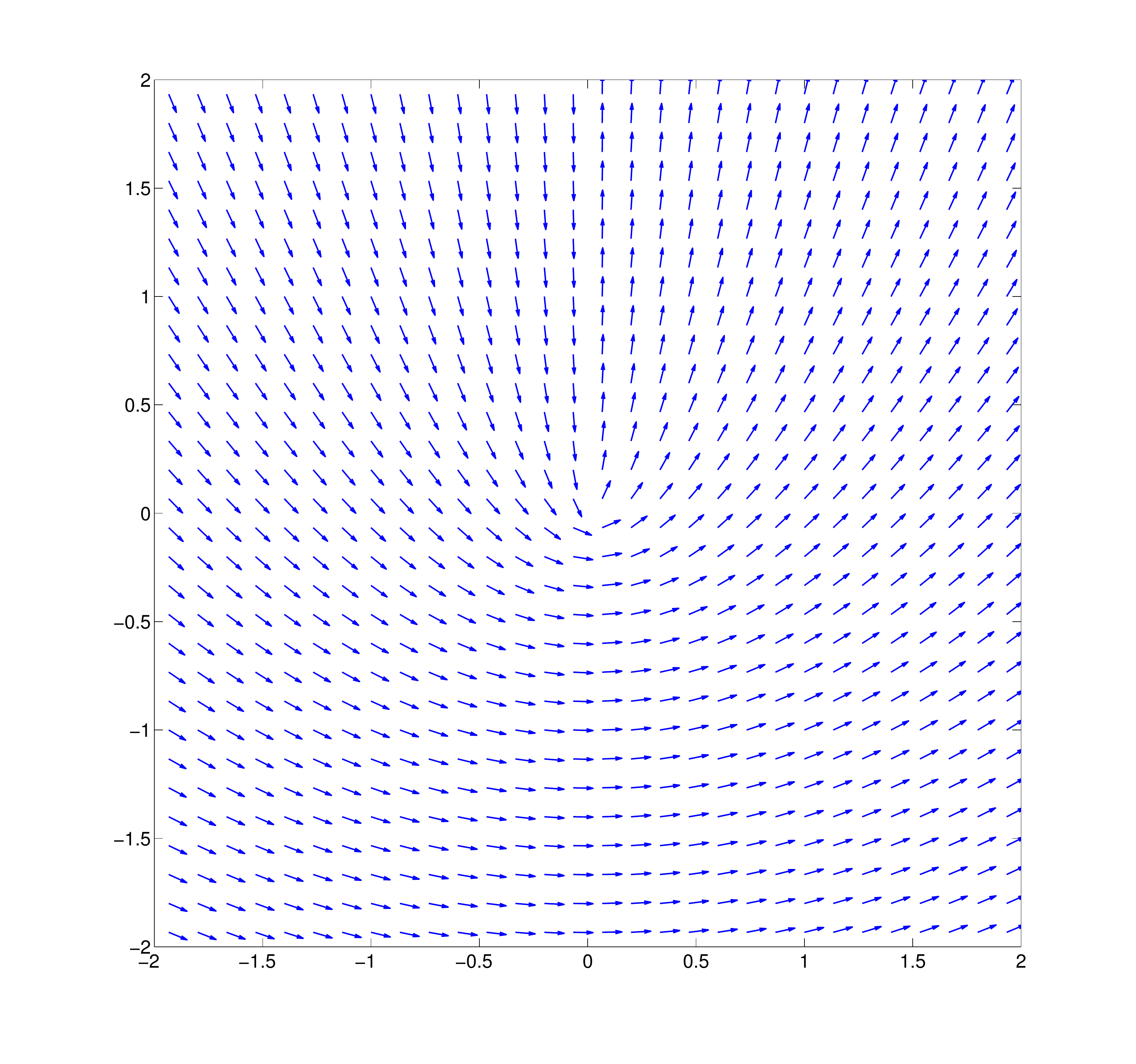}
	}
	\subfigure[$K=-\frac12$]{
		\includegraphics[width=50mm,trim=3.6cm 2.8cm 3cm 2cm, clip=true]{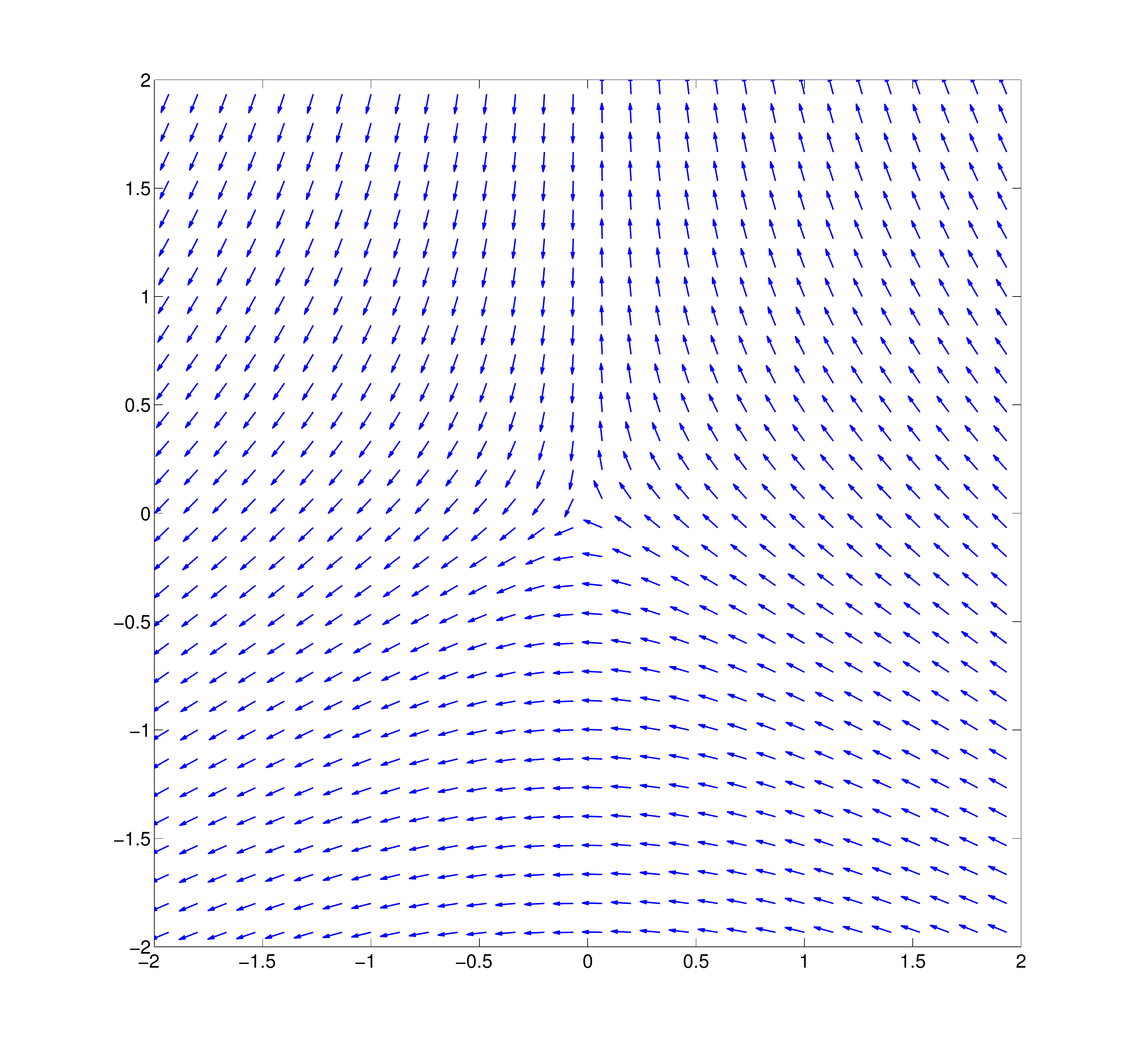}
	}
	\subfigure[$K=1$, $c=0$] {
		\includegraphics[width=50mm,trim=3.6cm 2.8cm 3cm 1.cm, clip=true]{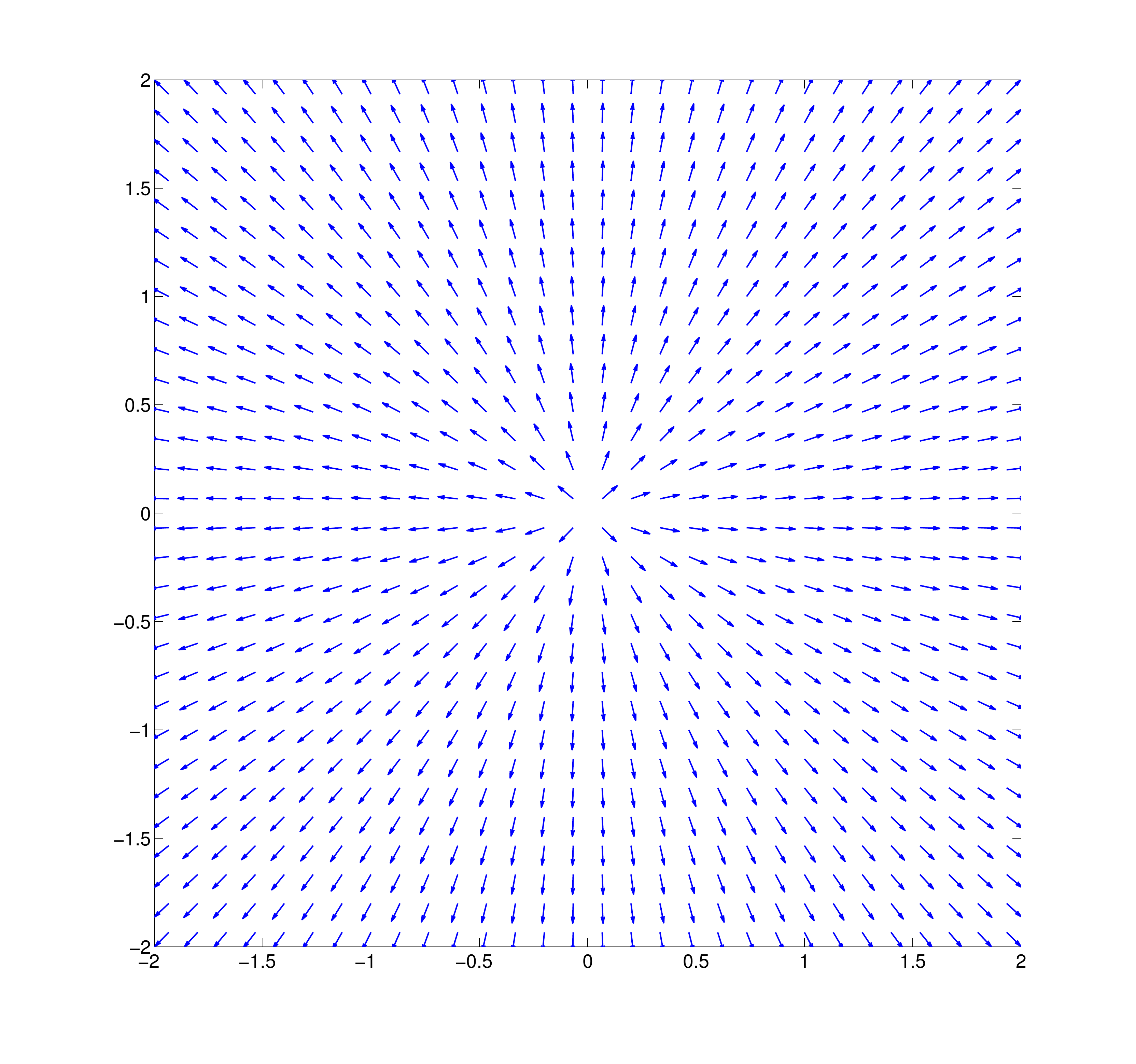}
	}
}
\centerline{
	\subfigure[$K=1$, $c=\pi/4$] {\label{k1_c4}
		\includegraphics[width=50mm,trim=3.6cm 2.8cm 3cm 2cm, clip=true]{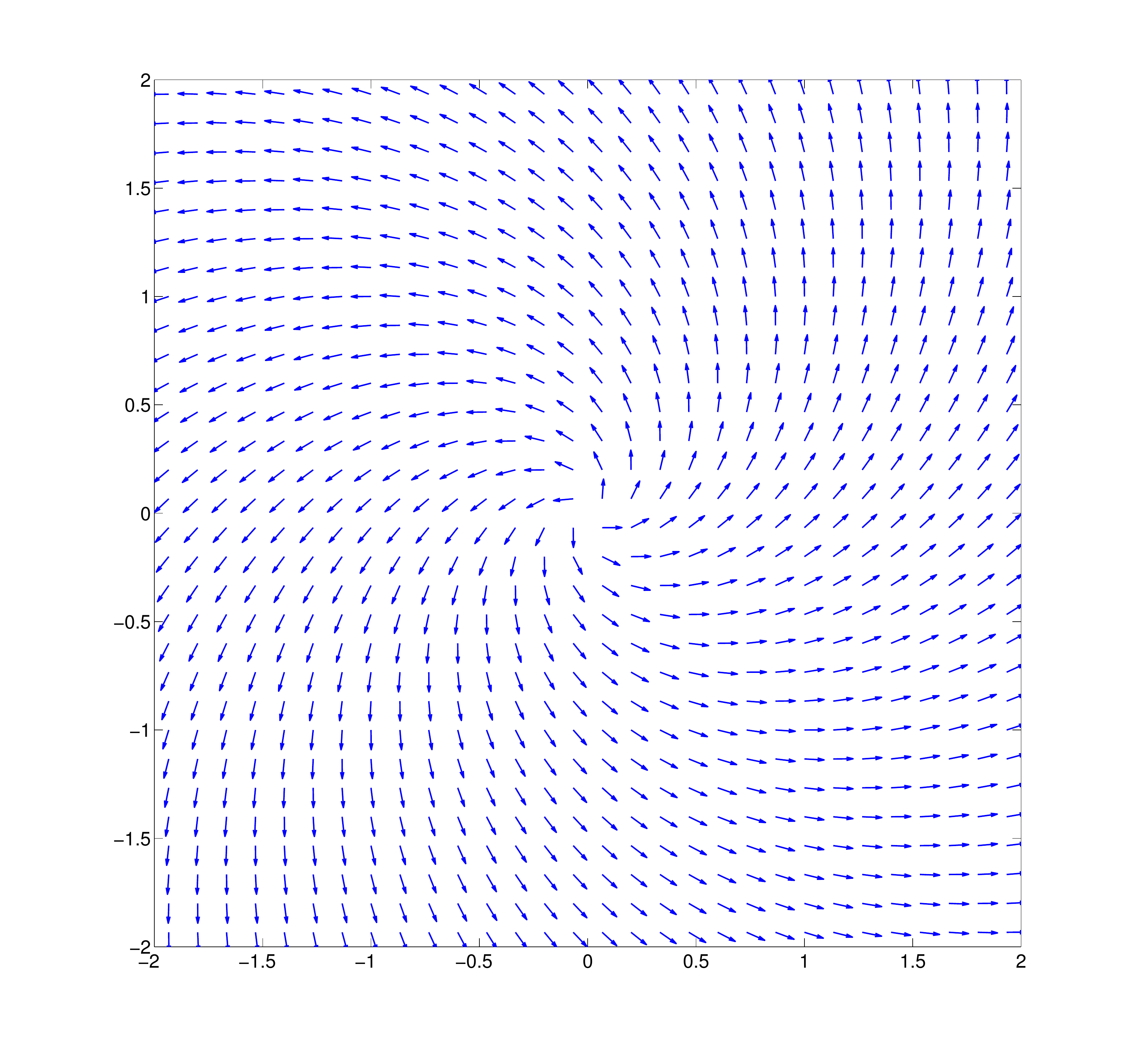}
	}
	\subfigure[$K=-1$]{
		\includegraphics[width=50mm,trim=3.6cm 2.8cm 3cm 2cm, clip=true]{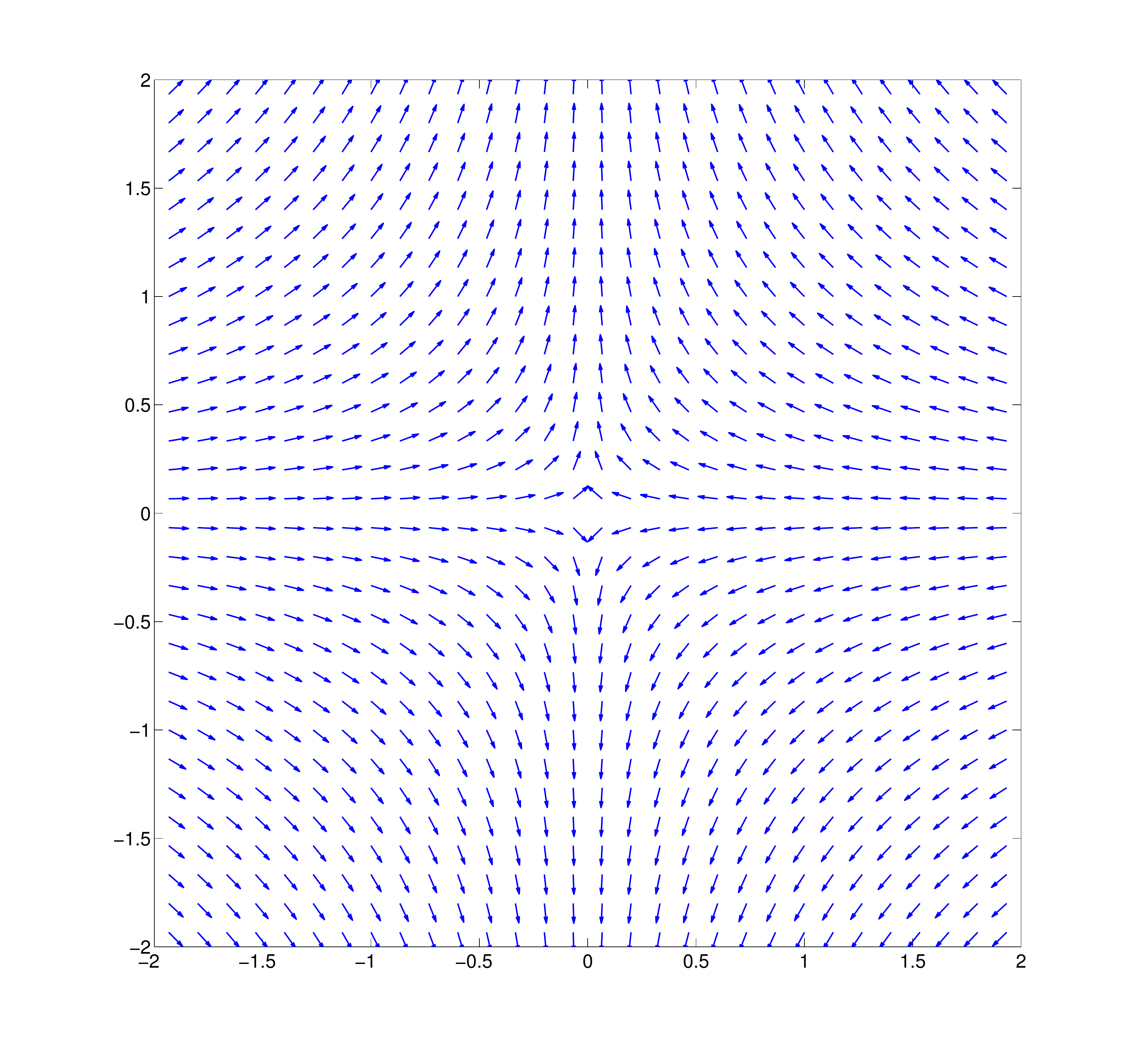}
	}
	\subfigure[$K=\frac32$]{
		\includegraphics[width=50mm,trim=3.6cm 2.8cm 3cm 2cm, clip=true]{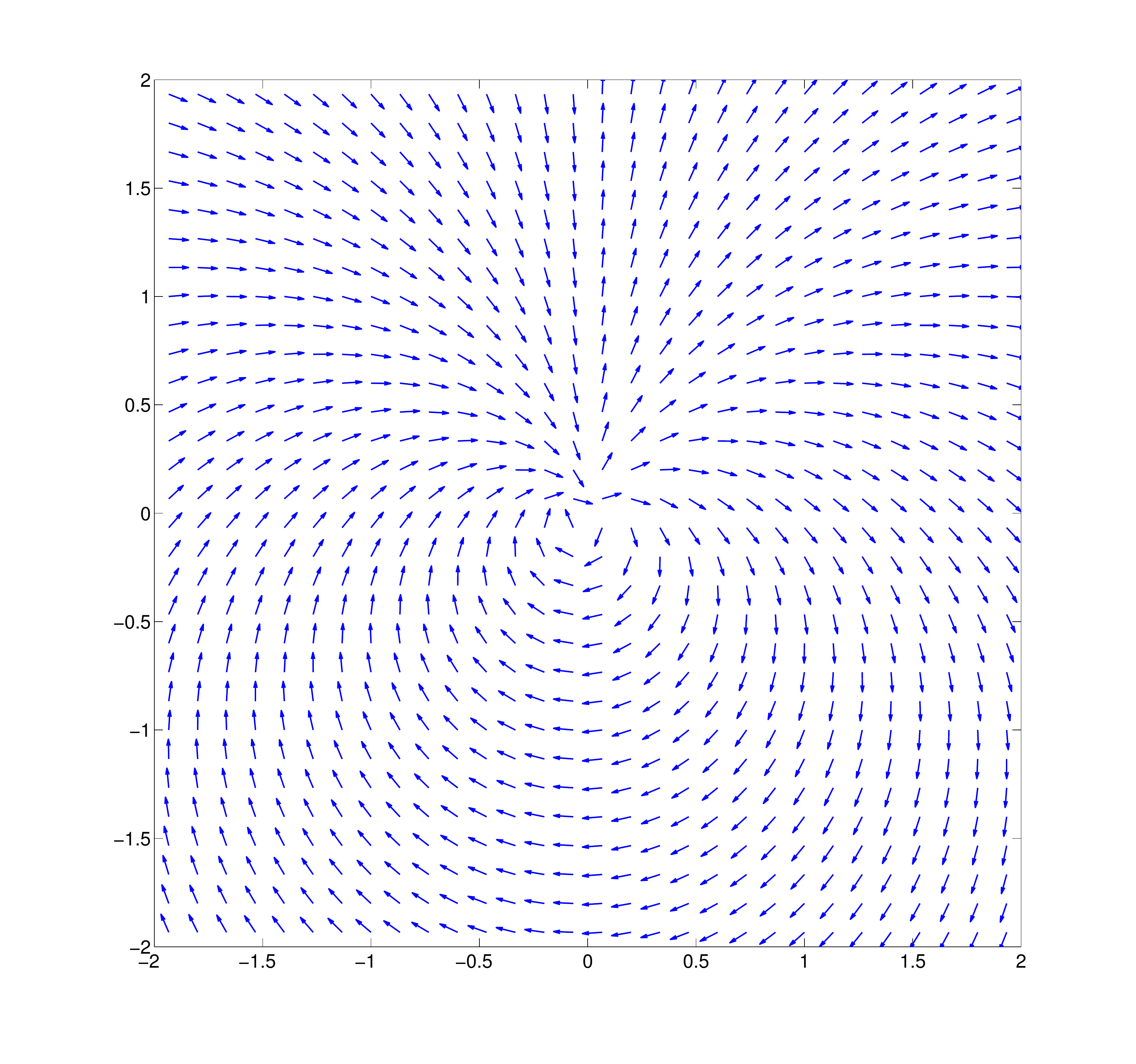}
	}
}
\centerline{
	\subfigure[$K=-\frac32$]{
		\includegraphics[width=50mm,trim=3.6cm 2.8cm 3cm 2cm, clip=true]{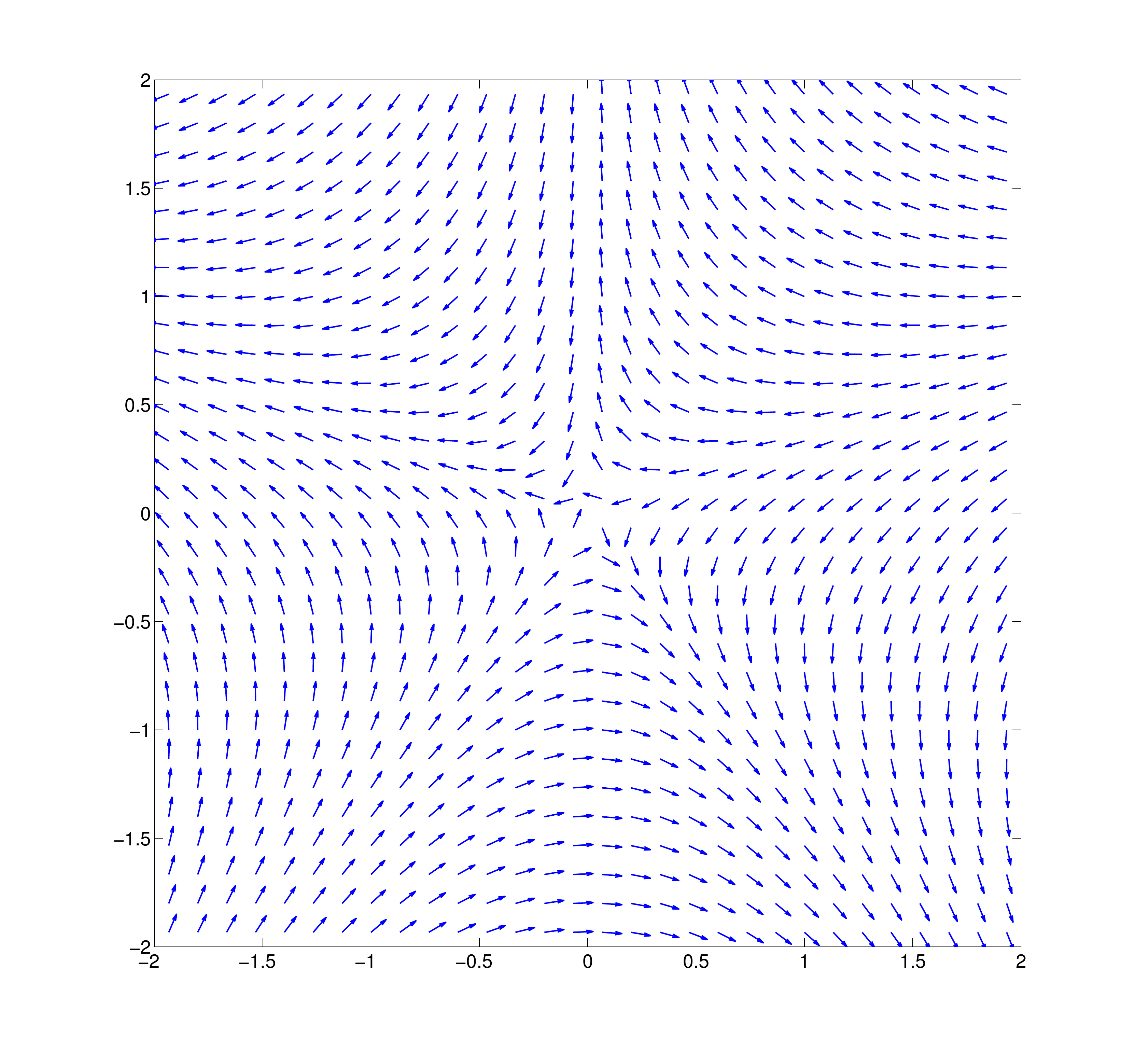}
	}
	\subfigure[$K=2$]{
		\includegraphics[width=50mm,trim=3.6cm 2.8cm 3cm 2cm, clip=true]{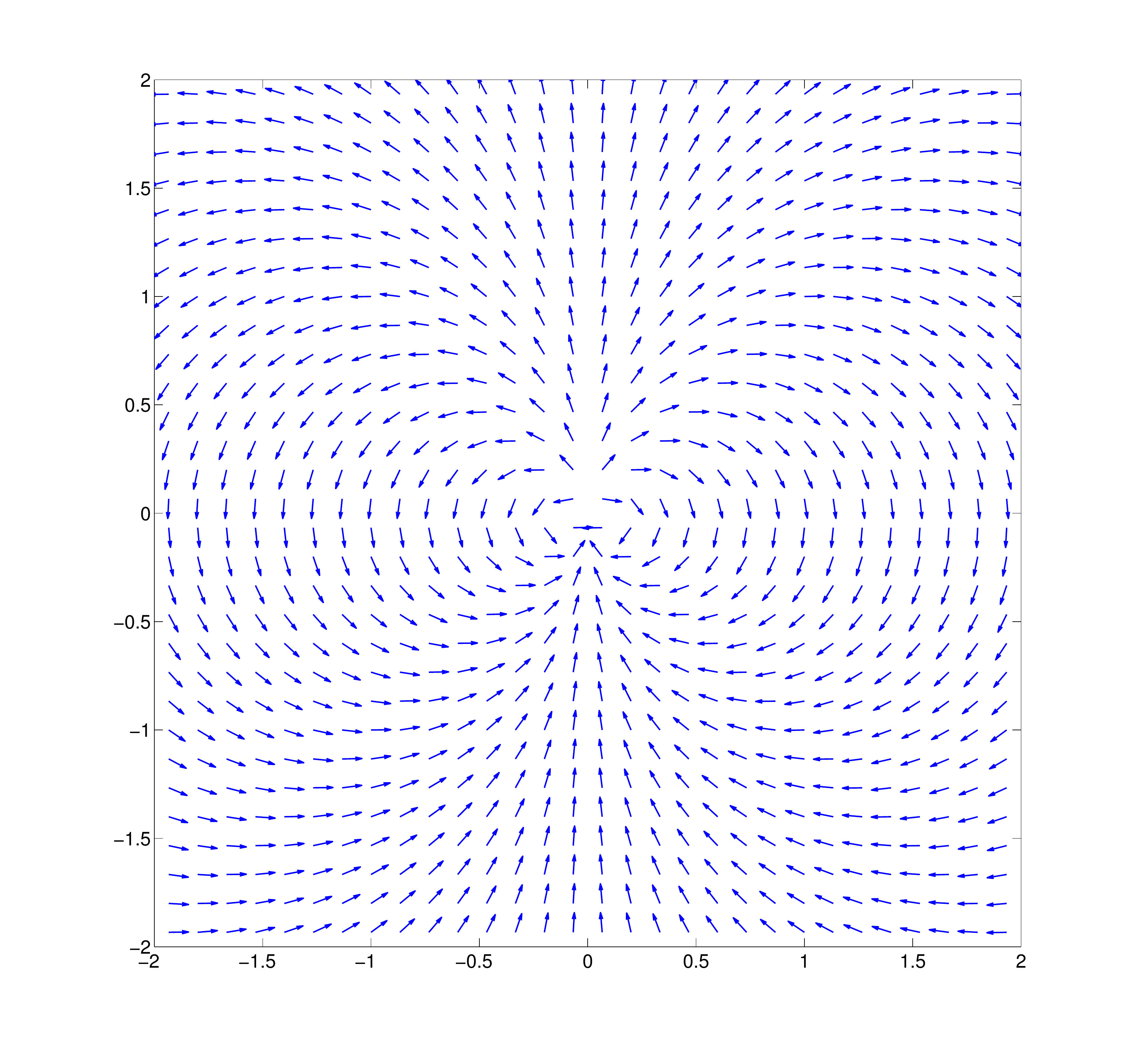}
	}
	\subfigure[$K=-2$]{
		\includegraphics[width=50mm,trim=3.6cm 2.8cm 3cm 2cm, clip=true]{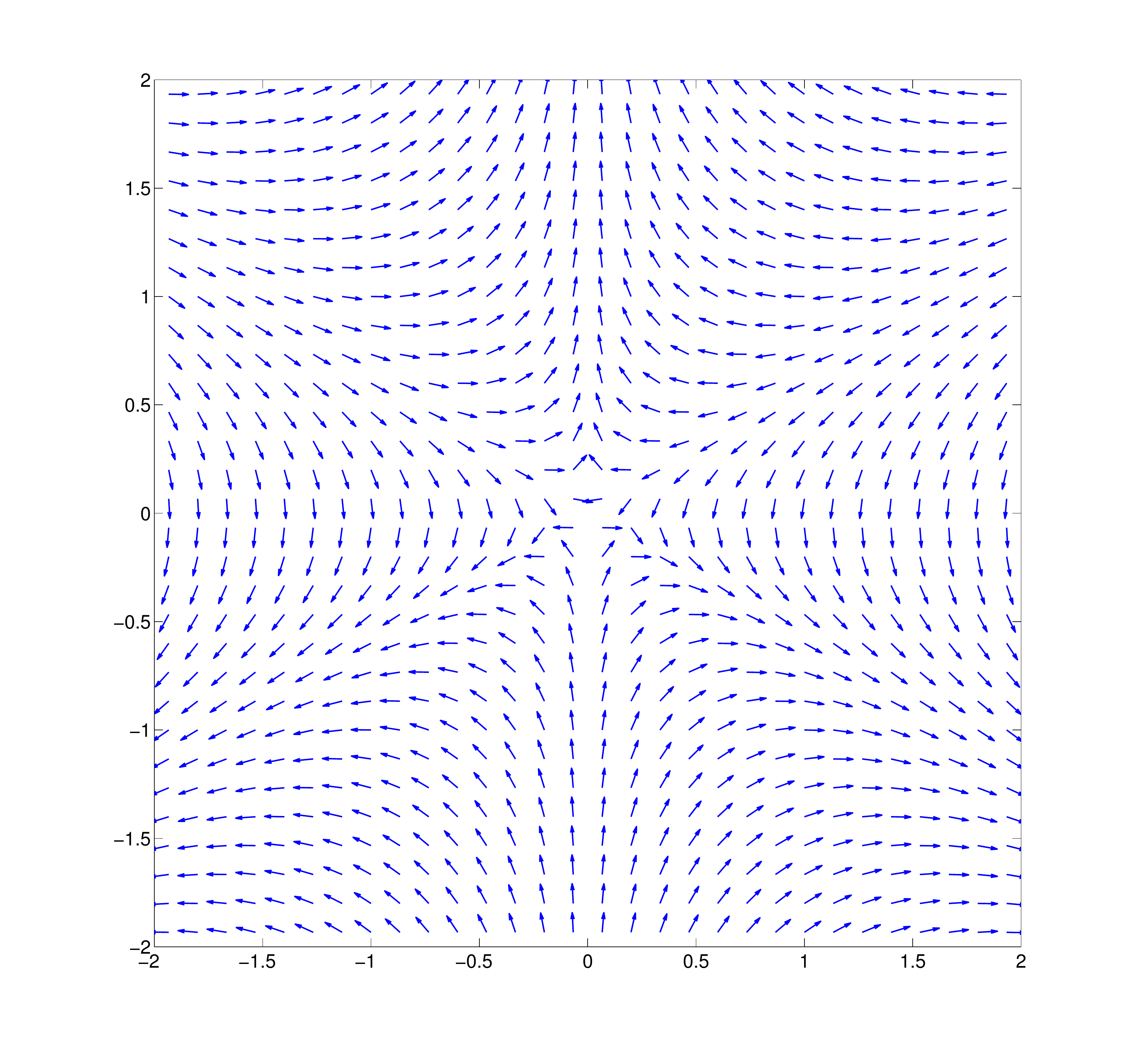}
	}
}
\caption{Computed director fields with wedge disclination at the center for different disclination strengths. Part (c) and (d) have the same disclination strength but different prescribed values $\theta_0$.}
\label{director}
\end{figure}

Figure \ref{director} shows the result of solving (\ref{weak-form}) for $\theta$, plotted as director fields using (\ref{n}), for different disclination strengths. In all of these examples, $\theta_0=\frac{\pi}2$, except in Figure \ref{k1_c4}. These director fields are completely analogous to those from Frank's solution (\ref{frank}), outside the layer.

\begin{figure}[!htb]
\centering{
	\includegraphics[width=46mm,trim=4.6cm 3.2cm 2.4cm 1.8cm, clip=true]{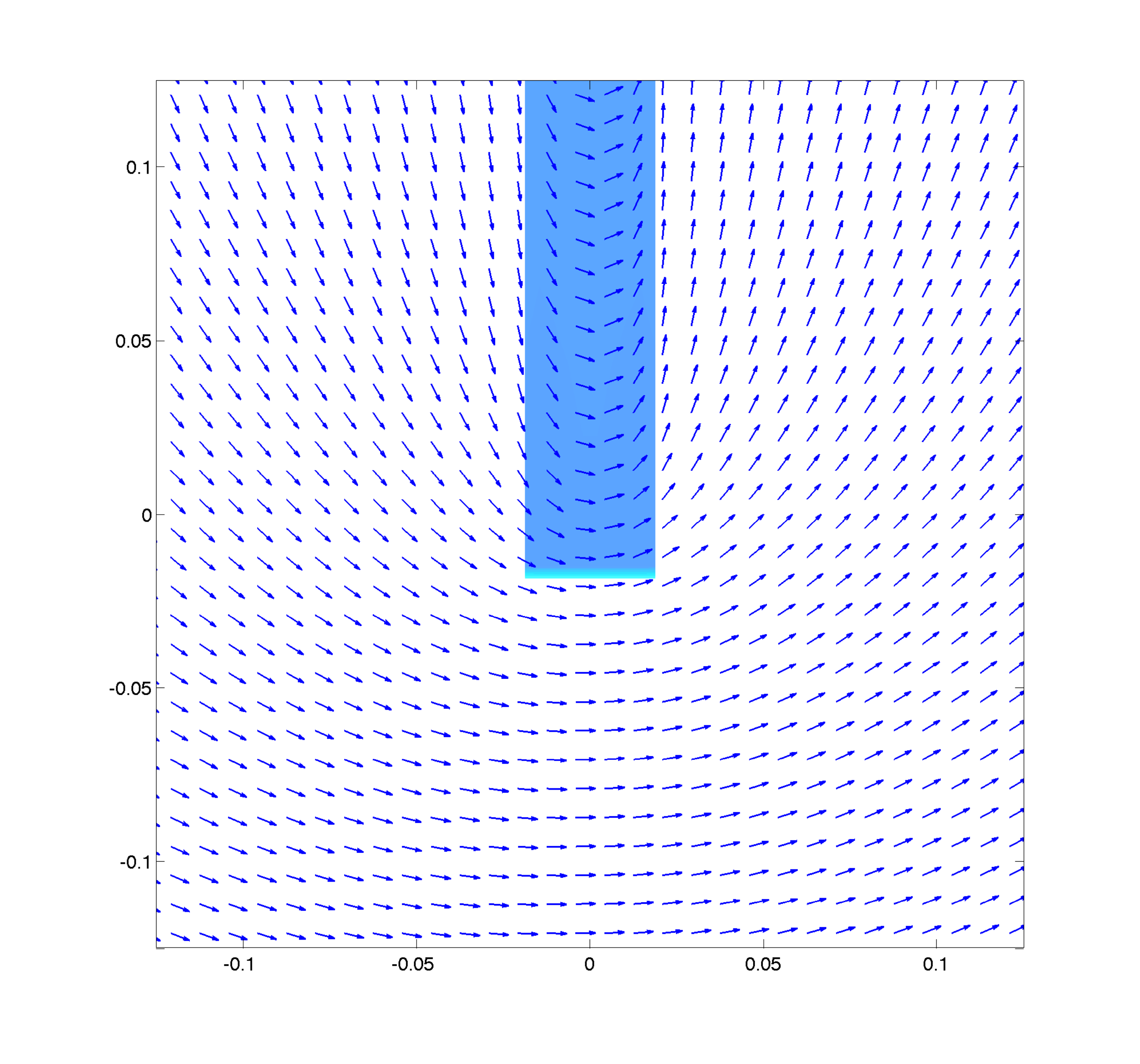}
}
\caption{A closer look at the computed director field for $K=\frac12$. The colored strip depicts the layer field, $\bflambda$. If one makes a clockwise circuit around the defect core, the rotation of the directors will be clockwise, outside of the layer, while it is anti-clockwise inside.}
\label{zoomed-director}
\end{figure}

However, inside the strip, the continuous $\theta$ field has to compensate the jump between the two sides of the layer. As a result, one can see that inside the layer, directors rotate sharply from one side to the other \emph{but at no extra energy cost in the layer beyond what is involved in the standard interpretation of the energy due to Frank's solution, i.e. the energy arising from considering the director gradient field to be (\ref{frank-grad})}. This is due to compensation by the $\bflambda$ field - more precisely, by the $grad\,z$ part of the $\bflambda$ field. Moreover, the topological content of the standard interpretation, i.e. (\ref{frank-grad}), of the singular gradient of Frank's solution (\ref{frank}) is preserved in the $\bfE^\theta$ field and approximately in the gradient of continuous $\theta$ field (up to the layer). Within this setting, even in the case of an integer defect, although there is no apparent discontinuity between the directors on the left and right side of the layer, there exists a $2\pi$ transition within the layer. In order to have a better understanding of this compensation, we take a closer look at the case $K=\frac12$. Figure \ref{zoomed-director} shows the rotation of the director field, inside the layer. Note that the rotation of the director field inside the layer is opposite in sense to its rotation outside of the field.

\begin{figure}[!htb]
\centering{
	\subfigure[] {
		\includegraphics[width=60mm,trim=2.cm 2.8cm 2.7cm 2.8cm, clip=true]{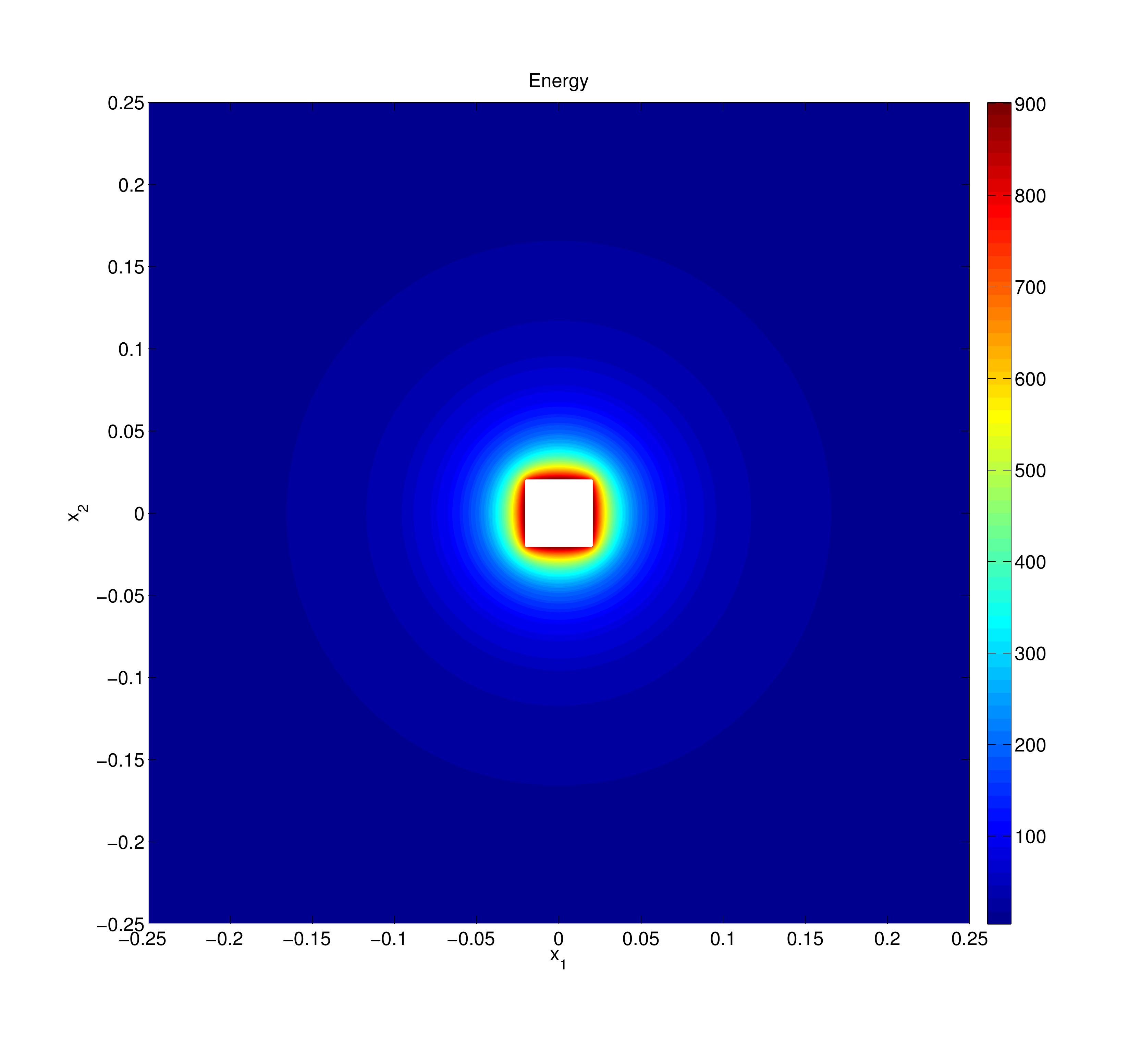}
	}
	\subfigure[] { \label{energy-decay}
		\includegraphics[width=70mm, trim=0.4cm 0.6cm .4cm 0.2cm, clip=true]{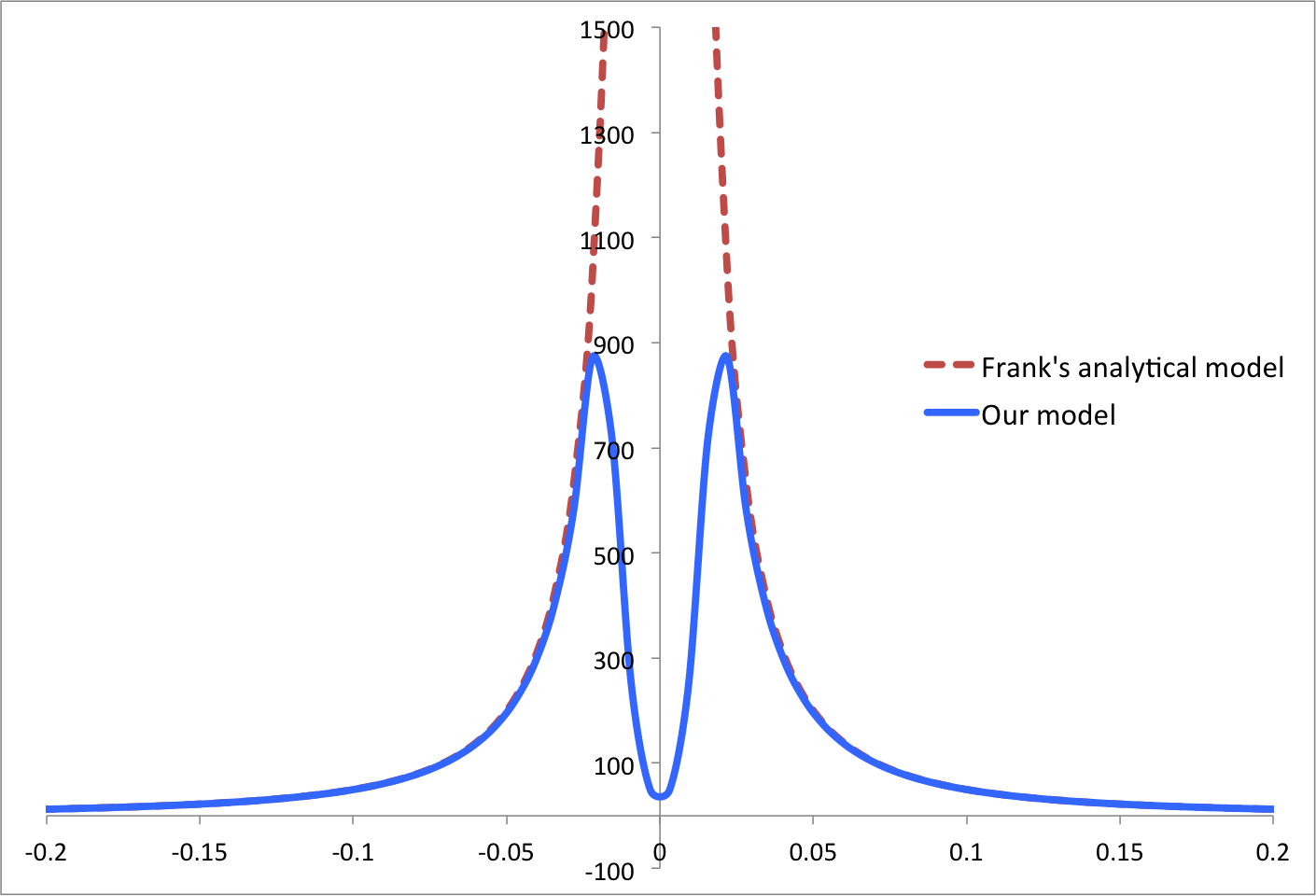}
	}
}
\caption{(a) Normalized energy density contour outside of the defect core and, (b) The decay of energy density when $x_2=0$.}
\label{energy-plot}
\end{figure}

Next, we compare the energy density field of our theory to the corresponding one of Frank's model. Using (\ref{1-const}), the energy term for Frank's planar model will be

\[ 
\psi_F(\bfn, \bfE)=\frac{\kappa}2\frac{K^2}{r^2}
\]
(using the standard interpretation of the gradient). Evidently, the total energy in a finite body for such an energy density field will not be finite. However, the energy density of the defects we model is non singular in most cases and definitely integrable in all cases, so that the total energy will be finite. The energy density of our model can be calculated from the following expression:

\[
\psi(\bfn, \bfE)=\frac12\kappa(\theta_{,i}-\lambda_i)(\theta_{,i}-\lambda_i).
\]
Figure \ref{energy-plot} shows energy contours normalized with respect to $K$ and $\kappa$, outside the defect core:
\[
\tilde{\psi_F} = \frac1{\kappa K^2}\psi_F=\frac1{2r^2} \ \ \ \ \ \tilde{\psi} = \frac1{\kappa K^2}\psi.
\]

In this Figure, we have also compared the decay of the energy density as it moves away from the defect core. According to $\psi_F$, such decay must follow a $\frac 1{r^2}$ trend. The results from our setting shows  good agreement with the analytical results outside the core, while being nonsingular inside, Figure \ref{energy-plot}.

\begin{figure}[!htb]
\centering{
	\subfigure[] {
		\includegraphics[width=60mm,trim=2.cm 2.6cm 2.8cm 2.8cm, clip=true]{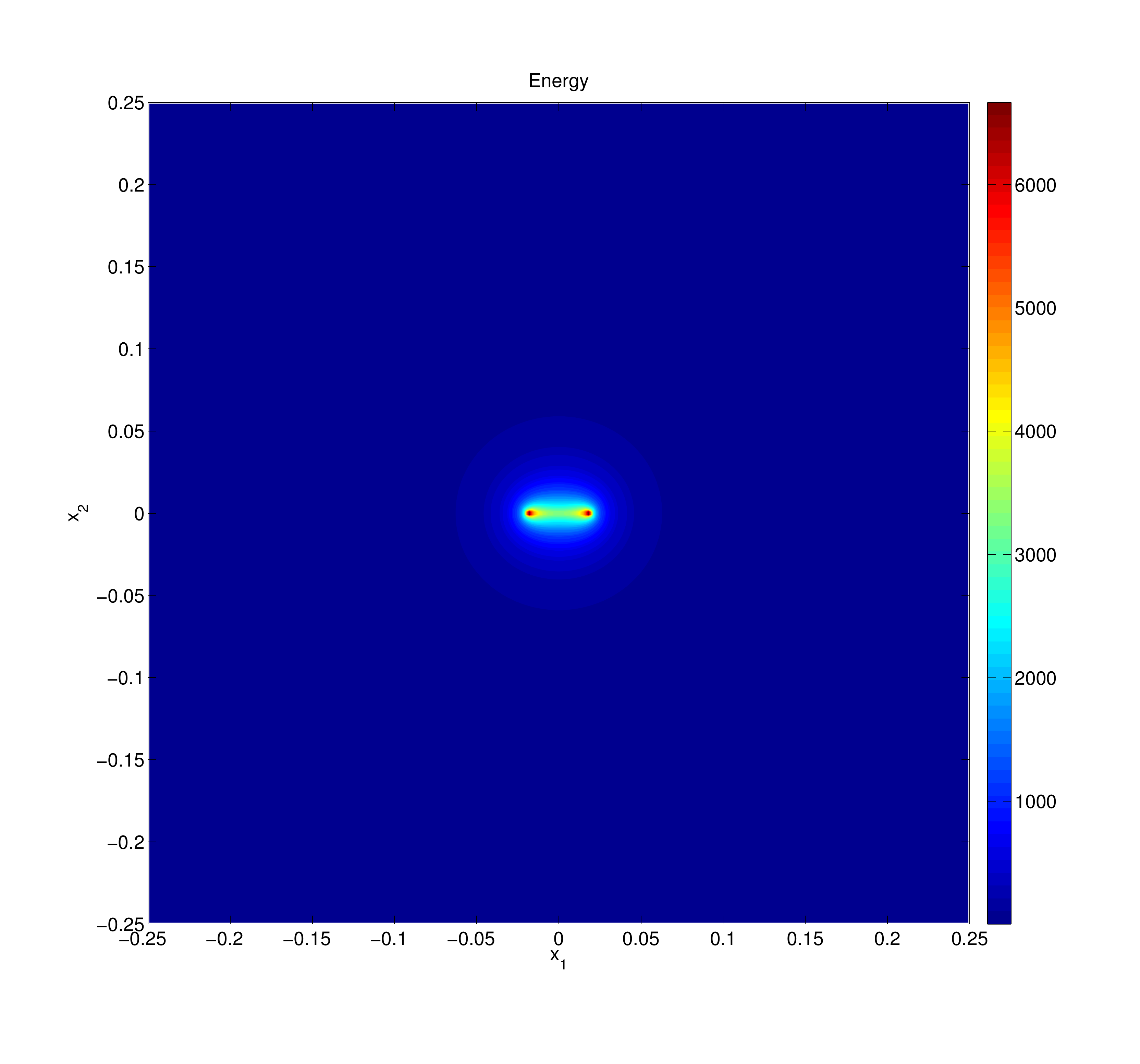}
	}
	\subfigure[] { \label{energy-flat-profile}
		\includegraphics[width=70mm, trim=0.4cm 0.3cm .4cm 0.2cm, clip=true]{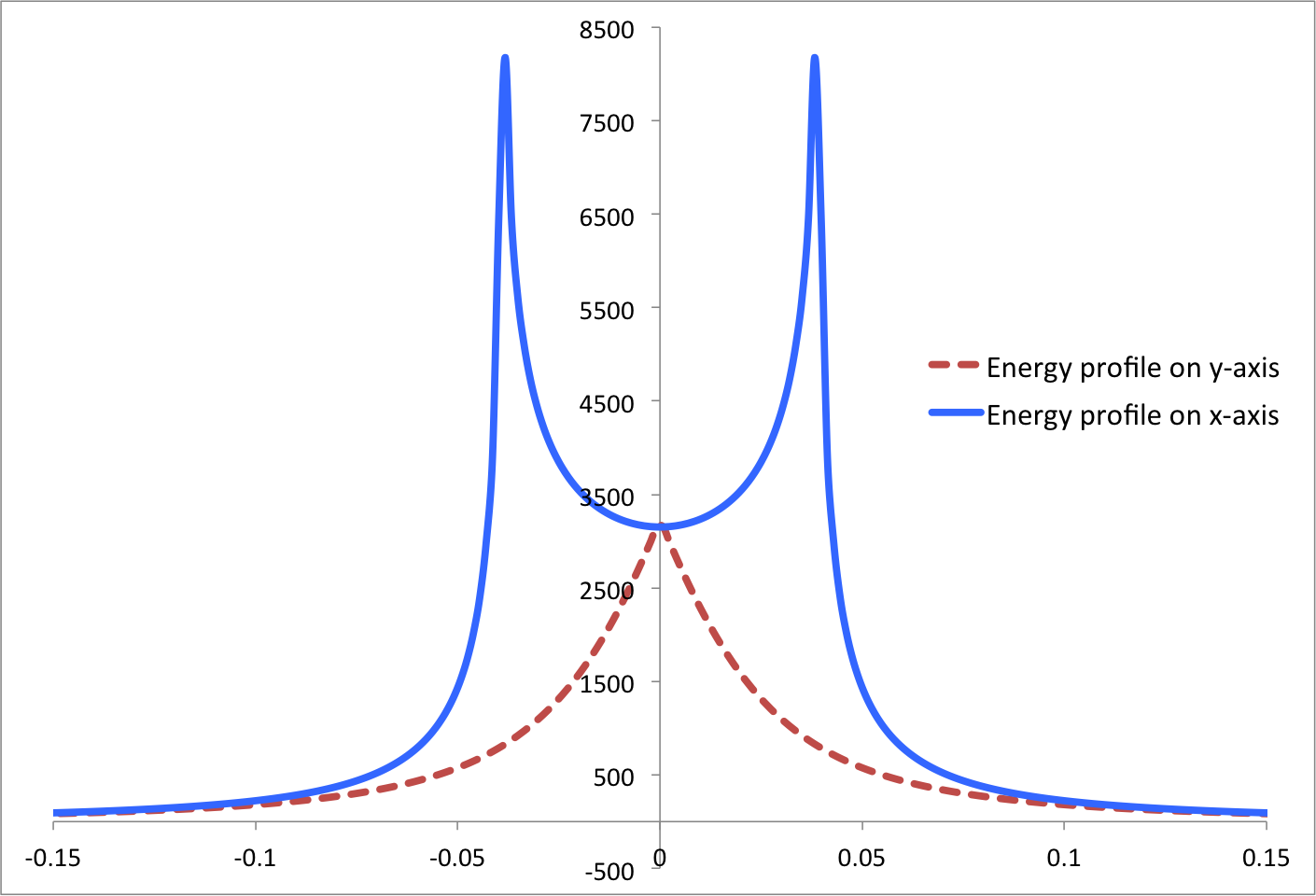}
	}
}
\caption{(a) Normalized energy density contour when $\bflambda$ sharply ends at the center, (b) The corresponding energy density profile in cases $x_2=0$ (blue, solid line) and $x_1=0$ (red, dashed line).}
\label{energy-flat-plot}
\end{figure}

Note that not only is the energy density in our model non-singular at the center, but Figure \ref{energy-decay} shows that the energy density profile drops dramatically at the center of the defect core. Later, in Figure \ref{energy-profile-conv} we will show that the energy level at this point converges to zero. However, this is not a general rule. In fact the energy density profile inside the core depends on the distribution of the layer field, $\bflambda$, and consequently $\bfbeta^{\theta}$. For instance, if we cut off the layer field at the center of the defect core instead of letting it decay linearly, i.e. use (\ref{sing_wall}) for the layer field definition, the energy density will be as in Figure \ref{energy-flat-plot}. Figure \ref{energy-flat-profile} reveals the fact that although such sharp cut off in the layer field will cause unsymmetric energy density distribution inside the defect core, the energy density will still be radially symmetric outside the core.
	
We also note that even though in this case we have a singular wall-like distribution of $\bfbeta^\theta$ at the core edge that can potentially cause singularities in the energy density at the sharp corners defining the layer, the total energy for a fixed width of layer will always converge as long as $r_1 > 0$, i.e. the layer has finite width. This is borne out in our computations where the total energy corresponding to Figure {\ref{energy-flat-plot} appears to converge, as shown in Figure \ref{total-energy-converge}.
\begin{figure}[!htb]
\centering{
	\includegraphics[width=70mm,trim=.2cm .5cm 0.1cm 0.2cm, clip=true]{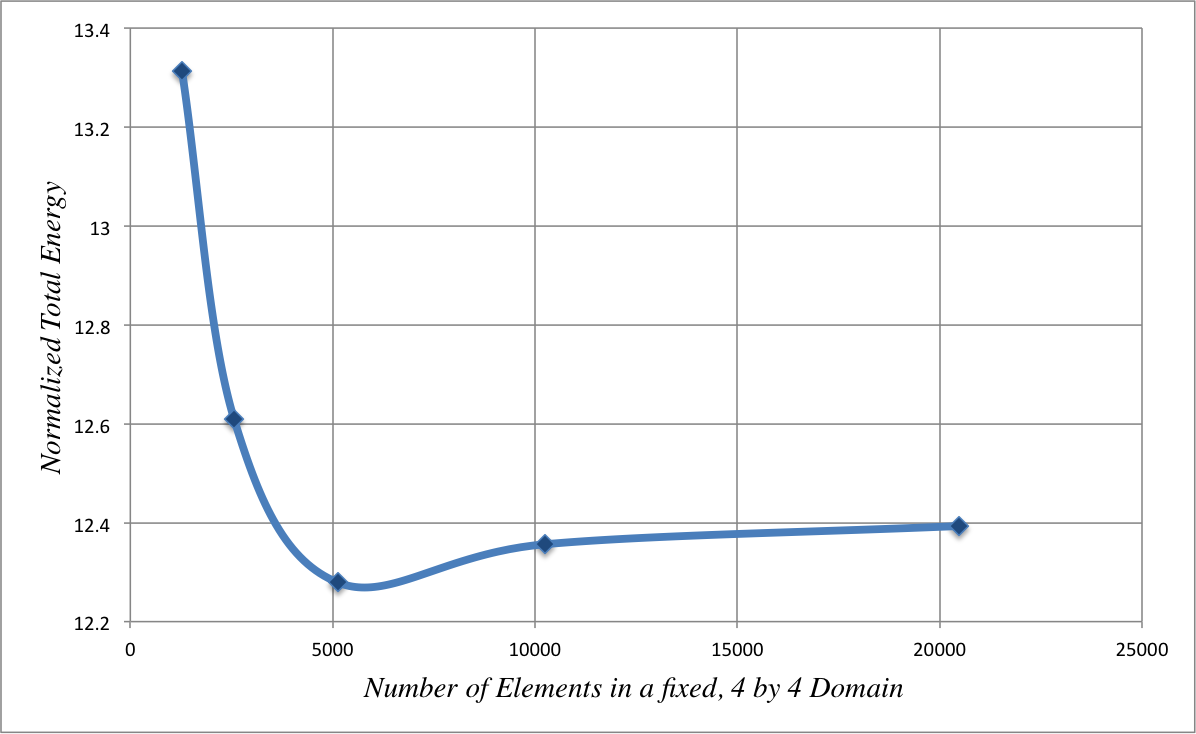}
}
\caption{Convergence of the normalized total energy, corresponding to the energy density shown in Figure \ref{energy-flat-plot} with the increase of the number of elements used in a fixed domain.}
\label{total-energy-converge}
\end{figure}
\subsubsection{Straight twist disclination}
Twist disclinations have the axis of rotation of the director field perpendicular to the defect line. They are sometimes called {\em perpendicular} disclinations. 

From the mathematical point of view, we use the same definition for the planar director field as in equation (\ref{n}). While the $\bfn$ field lies in the $x_1-x_2$ (horizontal) plane, $\theta$ is assumed to be a function of $x_1$ and $x_3$ and it is constant along the $x_2$ direction. The definition of the layer field undergoes similar, appropriate changes. We define $\bflambda$ as in (\ref{sing_wall}), but now in the  $\bfe_3$ direction instead of $\bfe_1$ (for direction of $\bflambda$ and the width of the layer) as shown in Figure \ref{straigth-twist-lambda}. We perform a 3-d finite element computation in this case. The computed energy density concentration is shown in Figure \ref{straigth-twist-energy}. The distribution of the energy and its decay on the vertical plane is exactly the same as shown in Figure \ref{energy-plot} for the straight wedge disclinations, which is consistent with expectation. Figures \ref{straigth-twist-director} and \ref{straigth-twist-director-anal}, respectively, show the computed and analytical director fields.

\begin{figure}[!htb]
\centering{
	\subfigure[] {
		\includegraphics[width=60mm,trim=1.6cm 1.2cm 2.3cm 2.2cmm, clip=true]{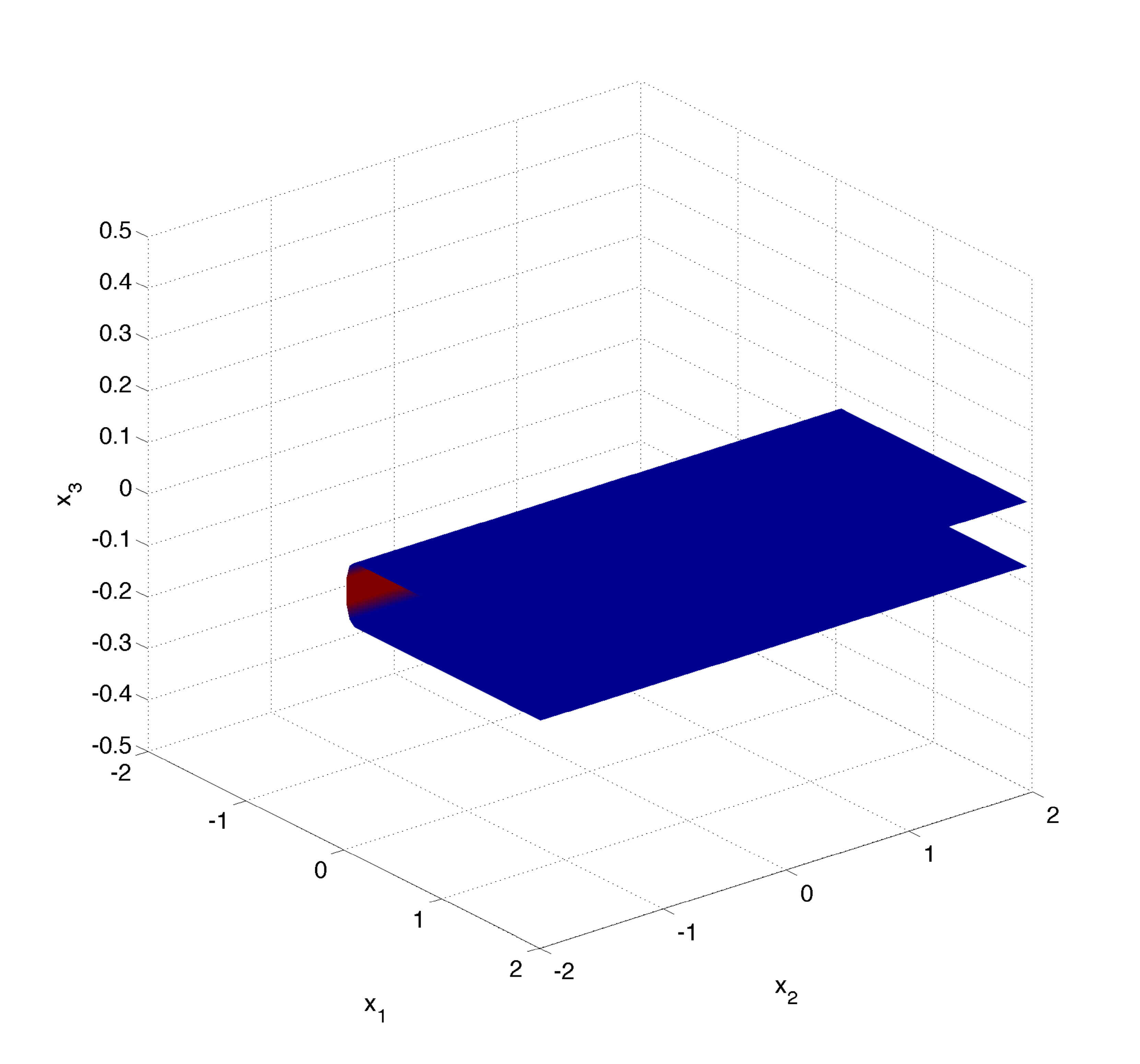}
		\label{straigth-twist-lambda}
	}
	\subfigure[] {
		\includegraphics[width=70mm,trim=2.3cm 0cm 2.6cm 1.8cm, clip=true]{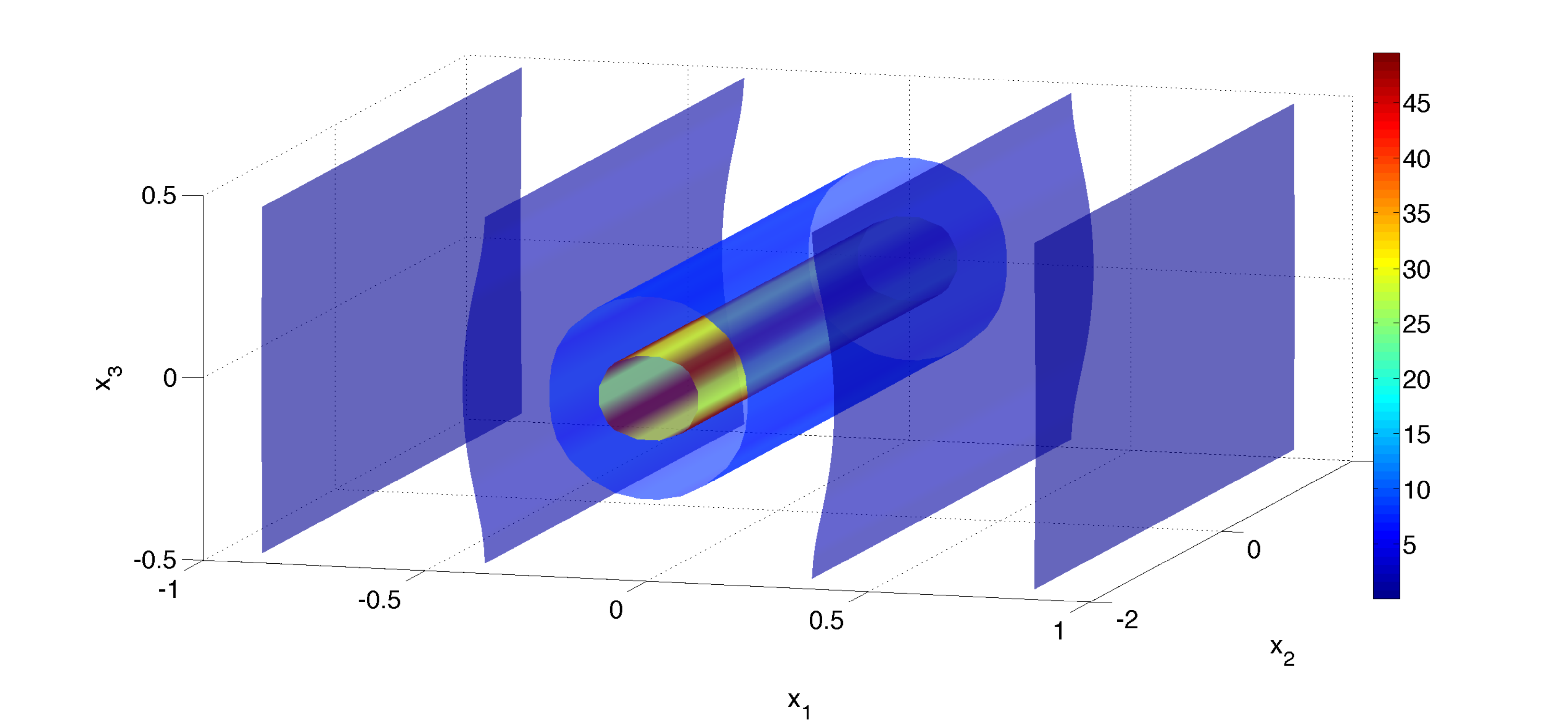}
		\label{straigth-twist-energy}
	}
	\subfigure[] {
		\includegraphics[width=65mm,trim=2.9cm .6cm 4.3cm 1.5cm, clip=true]{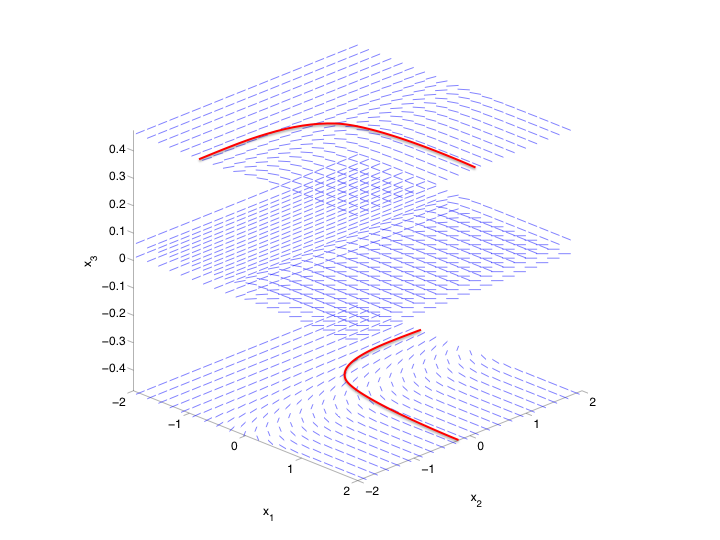}
		\label{straigth-twist-director}
	}
	\subfigure[] {
		\includegraphics[width=65mm, trim=1.5cm 1.cm 2.3cm 2.2cm, clip=true]{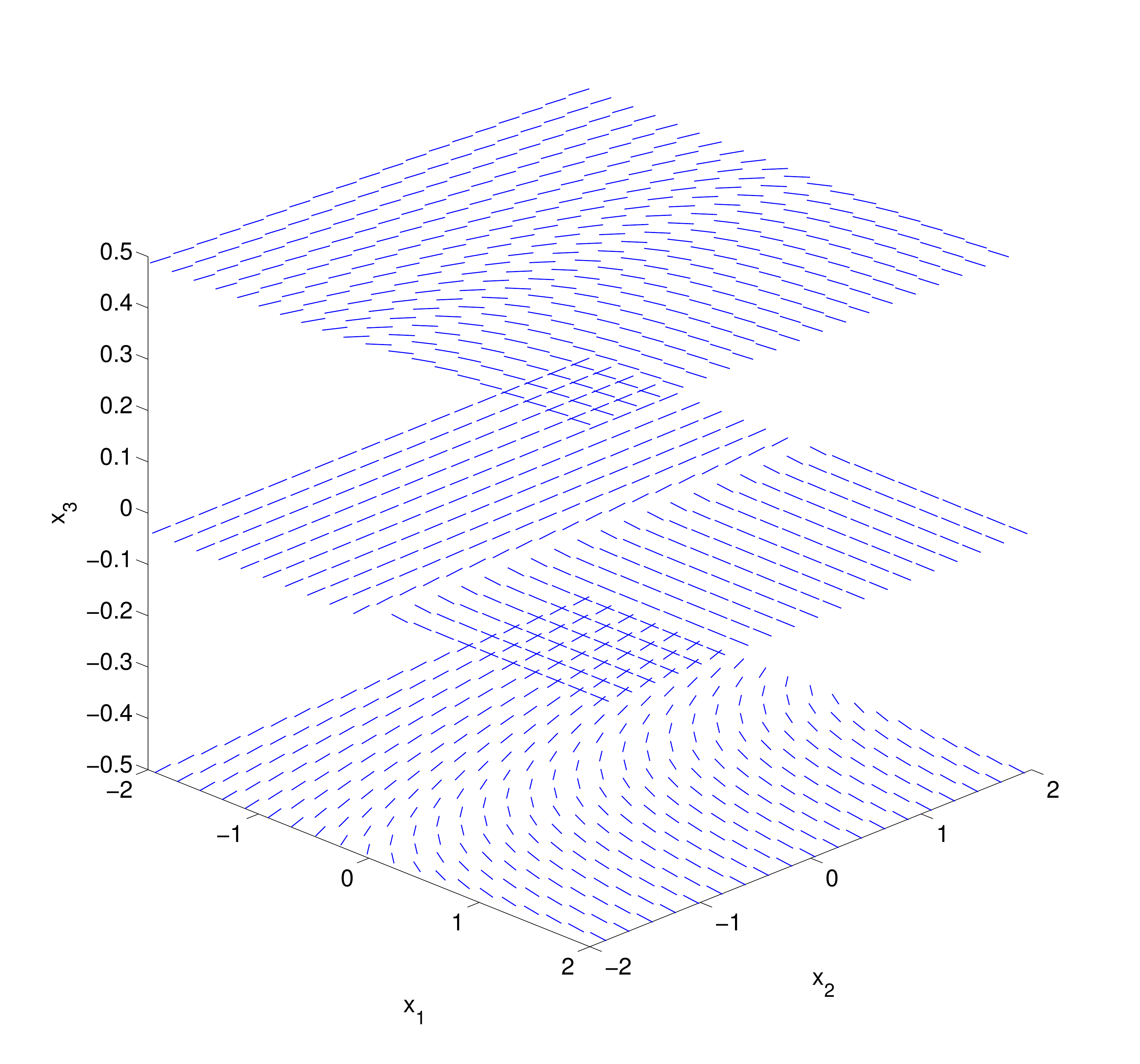}
		\label{straigth-twist-director-anal}
	}
}
\caption{Straight twist disclination: (a) 3D plot of the layer field, $\bflambda$, (b) a 3D plot of different layers of the normalized energy density to illustrate the energy density distribution in the body, (c) is the computed director field, while (d) is the analytical one.}
\label{straight-twist}
\end{figure}


\subsection{Twist Disclination loops}
Disclination loops are more commonly found in nature than curves terminating at boundaries \cite{Hwang:06}. In this section we will simulate a twist disinclination loop, using our framework.

The director field will be planar, but now the $\theta$ field will be a function of $x_1,x_2$ and $x_3$. The $\bflambda$ field for a planar twist disclination loop is taken to be a constant field in a bounded layer as can be seen in Figure \ref{lambda-loop}. The bounding cylinder of the layer over which the constant field decays to zero then forms the disclination loop. Clearly, the closed loop can be of any shape, but restricted to be planar in this exercise. 
\begin{figure}[h]
\centering{
	\includegraphics[width=60mm,trim=1.7cm 1.2cm 2.3cm 1.8cm, clip=true]{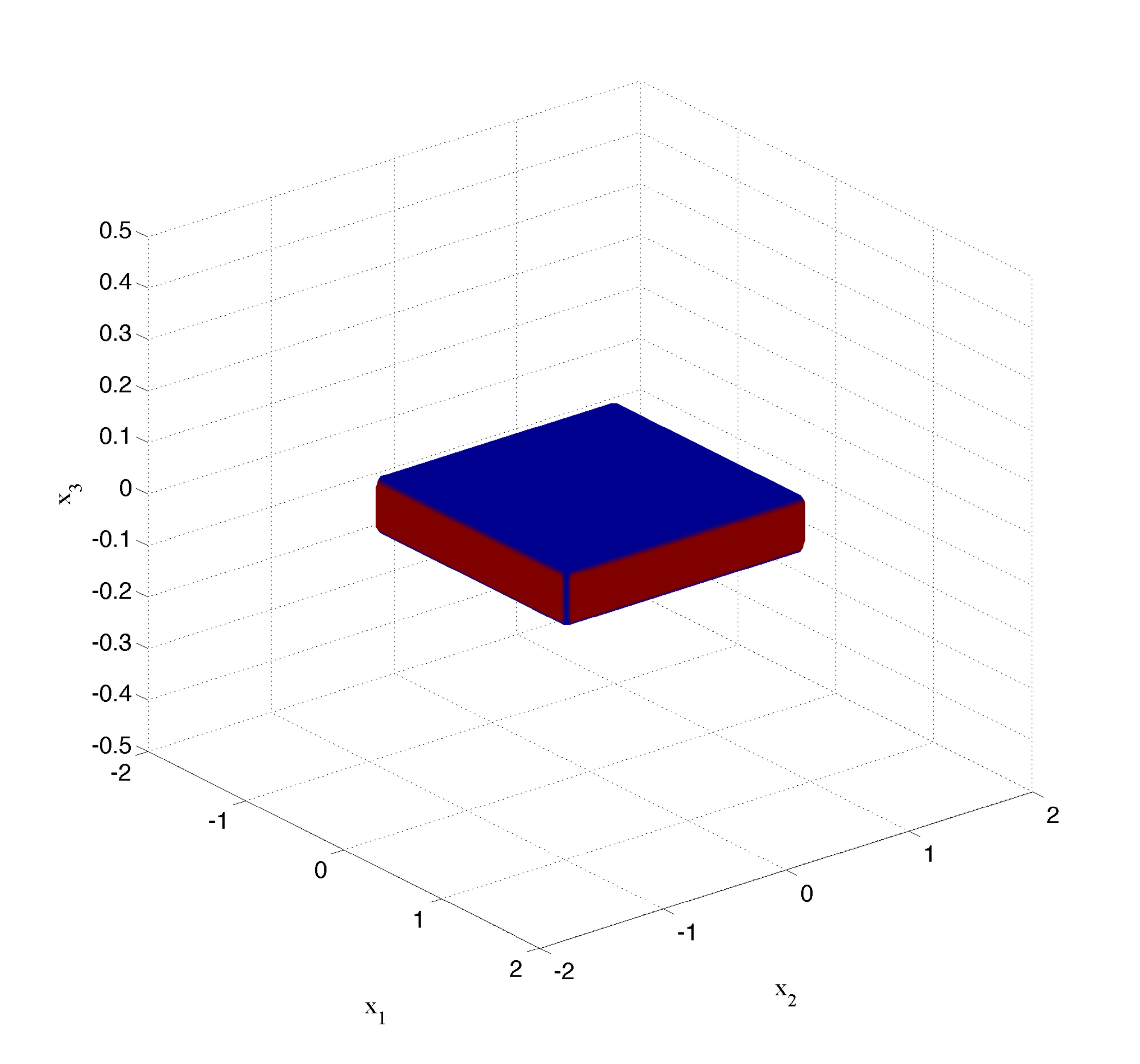}
}
\caption{An example of $\bflambda$ field for a square-shaped twist disclination loop. The field vanishes in the white region.}
\label{lambda-loop}
\end{figure}

In particular, we have chosen a square-shaped and a circular loop in the horizntal plane.  $\bfbeta$ and $\bfbeta^{\theta}$ field for the rectangular loop are shown in Figure \ref{beta-loop}. Just like the 2-dimensional wedge disclination case, the $\bfbeta^{\theta}$ field is non-zero only at the defect core which appears as a square loop, and $\bfbeta$ is non-zero inside the loop, where the layer field is non-zero.

\begin{figure}[!htb]
\centering{
	\subfigure[] {
		\includegraphics[width=60mm,trim=2.cm 1.1cm 2.5cm 1.8cm, clip=true]{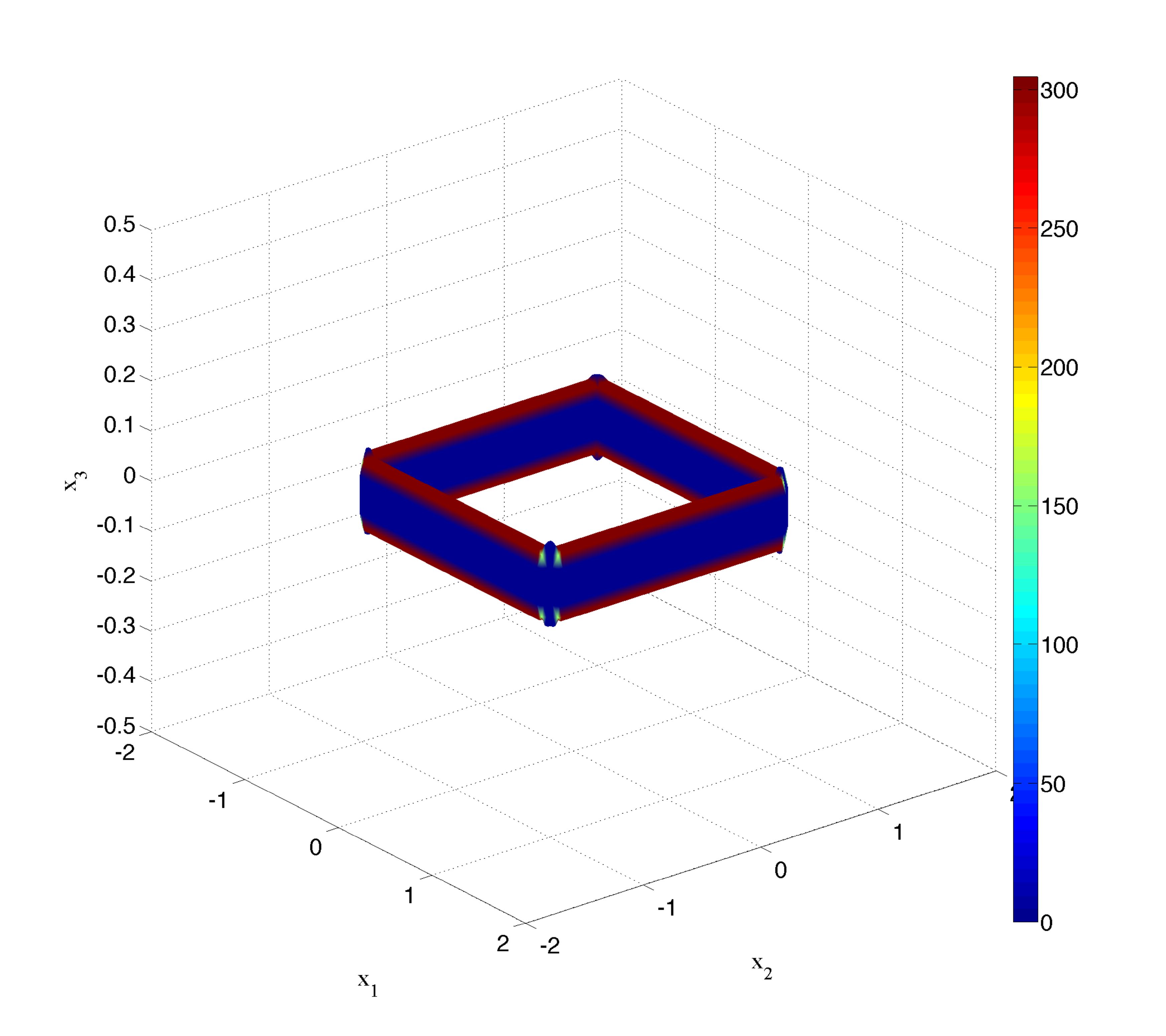}
	}
	\subfigure[] {
		\includegraphics[width=60mm,trim=2.3cm 2.5cm 2.5cm 2.7cm, clip=true]{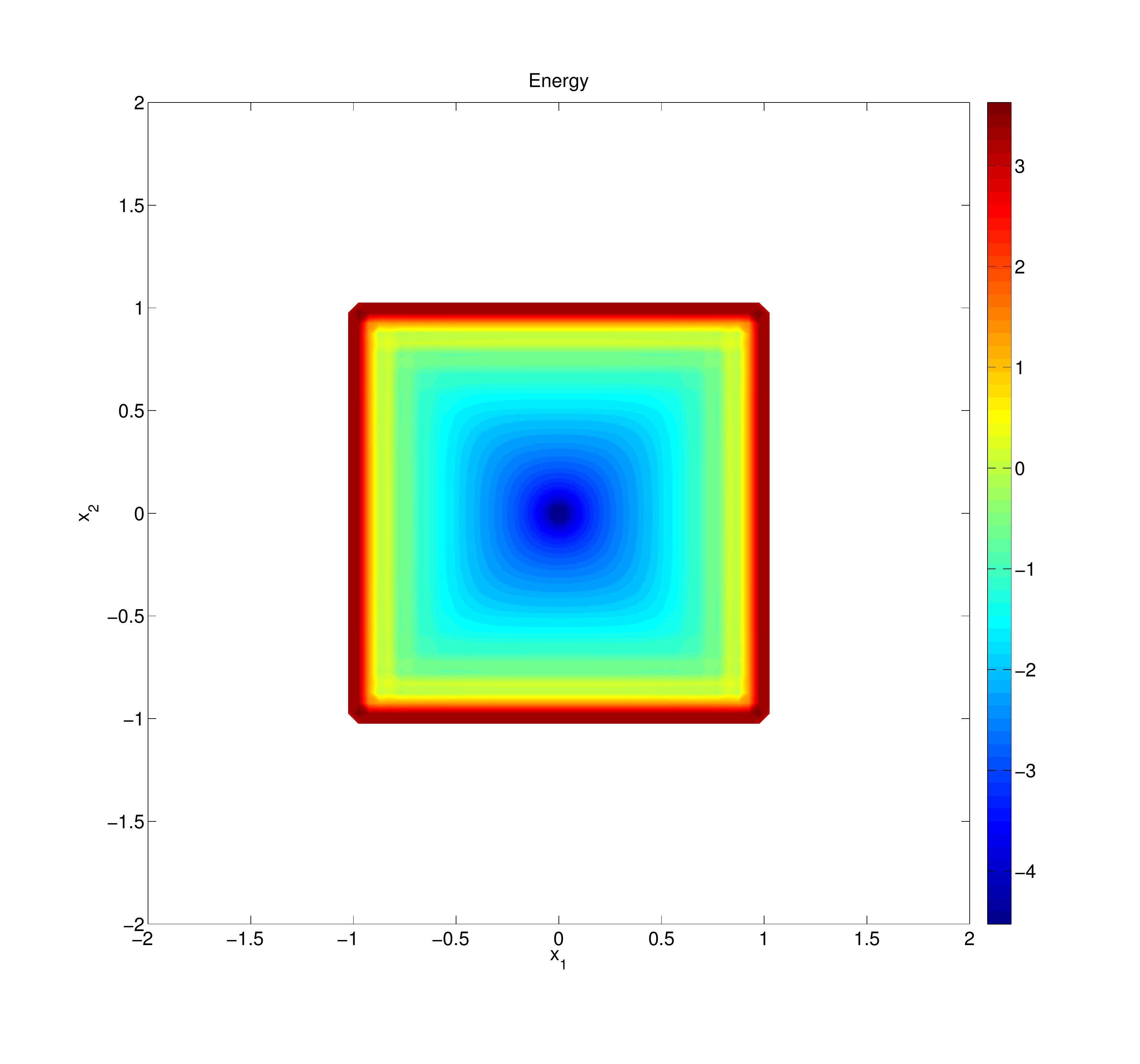}
	}
}
\caption{(a) Normalized $\bfbeta^{\theta}$ for the twist disclination loop and (b) Logarithmic contour plot of normalized $|\bfbeta|$.  The white area in part (b) represents zero for $|\bfbeta|$.}
\label{beta-loop}
\end{figure}

The director field is a planar vector field with twist axis along the $x_3$ direction. As shown in Figure \ref{director-loop}, in any given plane $x_3 = constant$ the effect of the loop decays rapidly with distance from the loop. Figure \ref{director-loop} also clarifies the way in which the director vectors make a twist as one moves through a circuit enclosing the core. Note that the director vectors corresponding to the lower part of the $\bflambda$ field have the opposite direction of the ones on the top, since that is a half-integer disclination loop ($K=\frac12$) and as a result the director field makes a $\pi$-radian twist. We emphasize that while this twist deficit is observable in the (continuous) director field only on circuits that are `closed up to the thickness  of the core', integrating $\bfE^\theta$ on an exactly closed circuit encircling an arm of the disclination loop core would yield a $\pi$-radian twist by design.

\begin{figure}[!htb]
\centering{
	\subfigure[Square-shaped loop] {
		\includegraphics[width=65mm,trim=3.7cm 1.2cm 3.5cm 1.4cm, clip=true]{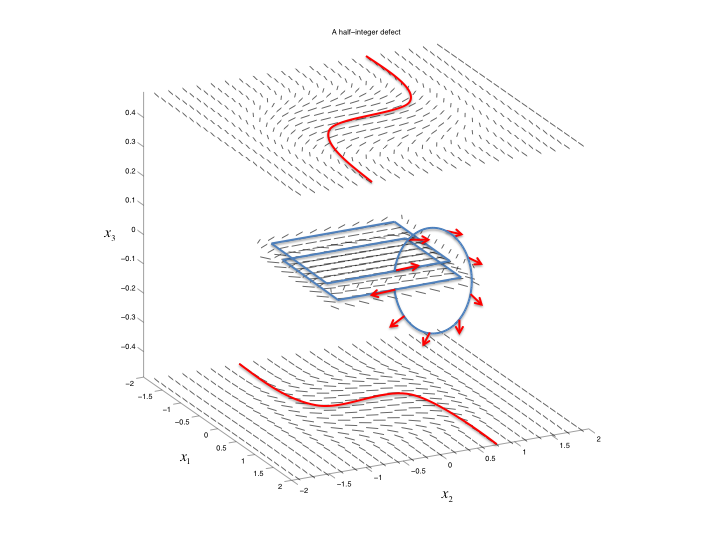}
	}
	\subfigure[Circular loop] {
		\includegraphics[width=65mm,trim=3.4cm 1.2cm 3.8cm 1.4cm, clip=true]{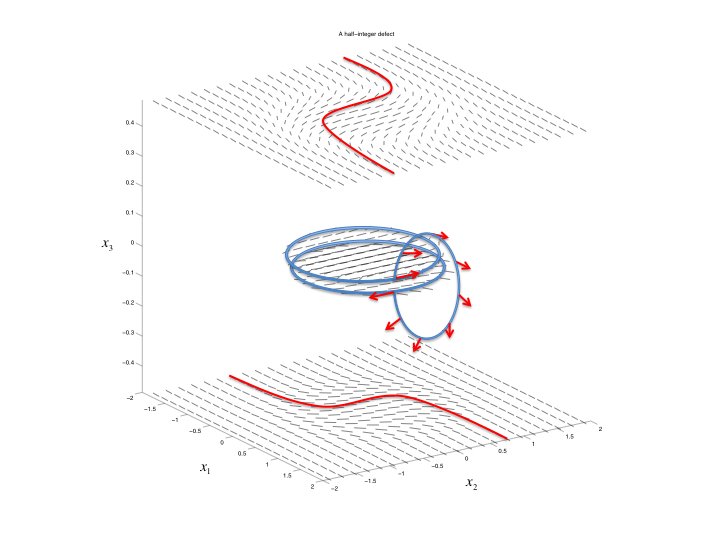}
	}
}
\caption{Computed director fields of twist disclination loops for $K=\frac12$.}
\label{director-loop}
\end{figure}

Figure \ref{energy-loop} shows the energy plots of the two disclination loops from different points of view. Figures \ref{energy-xy} and \ref{energy-xy-circ} are the energy density contours at $x_3=0$, on the $x_1-x_2$ plane for the square-shaped and circular loop, respectively. Note that the energy density of the rectangular loop has significantly higher magnitude at the corners, this being a consequence of the high curvature in the director field at the corners of the loop. Similar to what we saw in Figure \ref{energy-decay}, these contour plots show a drop in magnitude inside the core. 
This plot also shows smoothness of the energy density, even in the vicinity of the defect core. In order to better understand the pattern of the energy density on this plane, Figures \ref{log-energy-xy} and \ref{log-energy-xy-circ} show the logarithmic scale of the plot. A closer look at these two plots reveals that even for the rectangular defect core the energy density adopts a radially symmetric patter away from the core. The observable deviation from this circular texture on the boundary is due to the zero Neumann boundary condition that we have used for these two examples. Figures \ref{log-energy-xz} and \ref{log-energy-xz-circ} show the logarithmic plot of the energy density on the $x_1-x_3$ plane. 

\begin{figure}[!htb]
\centering{
	\subfigure[] {\label{energy-xy}
		\includegraphics[width=60mm,trim=2.3cm 2.5cm 2.5cm 2.8cm, clip=true]{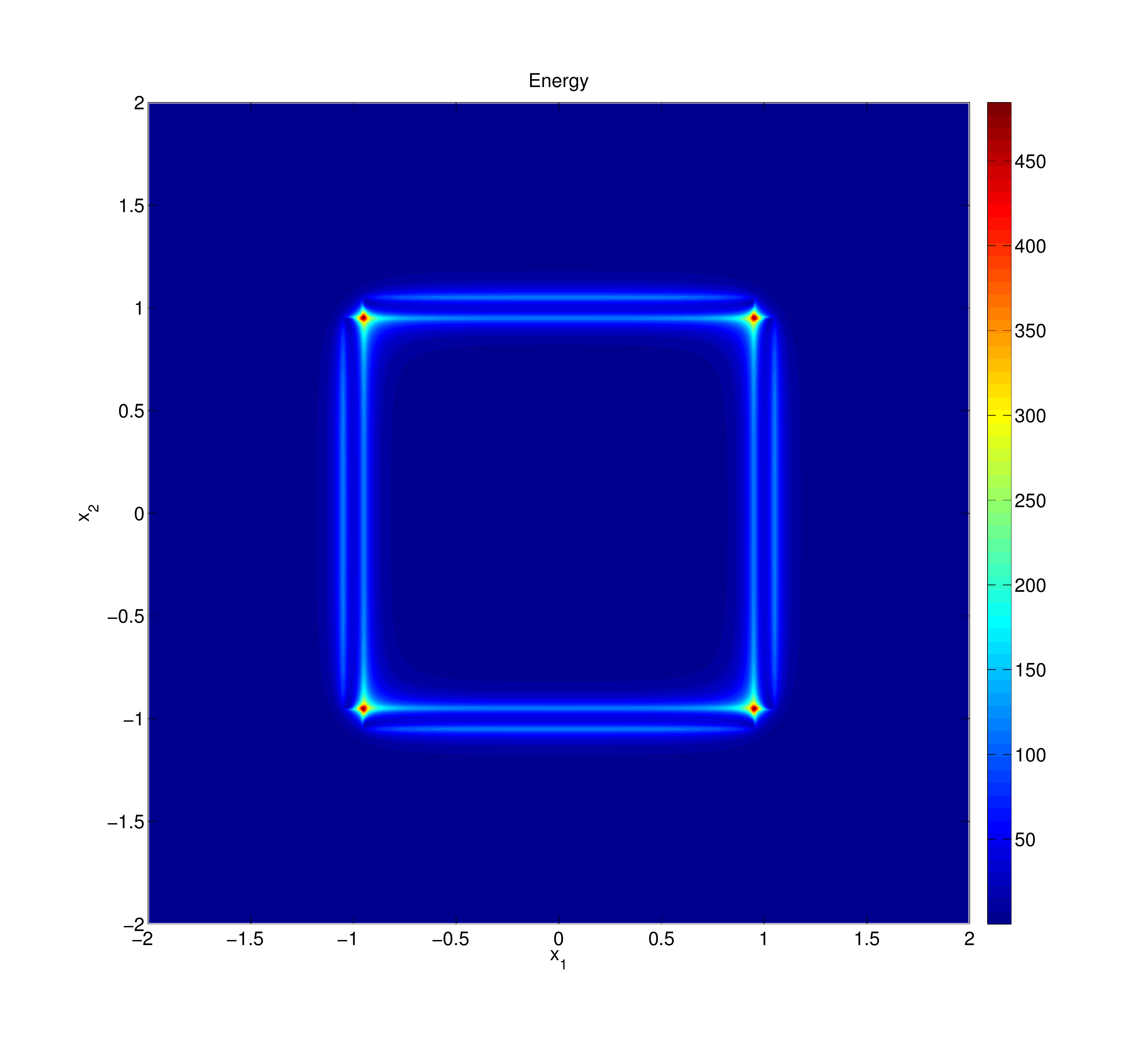}
	}
	\subfigure[] {\label{energy-xy-circ}
		\includegraphics[width=60mm,trim=2.3cm 2.5cm 2.3cm 2.8cm, clip=true]{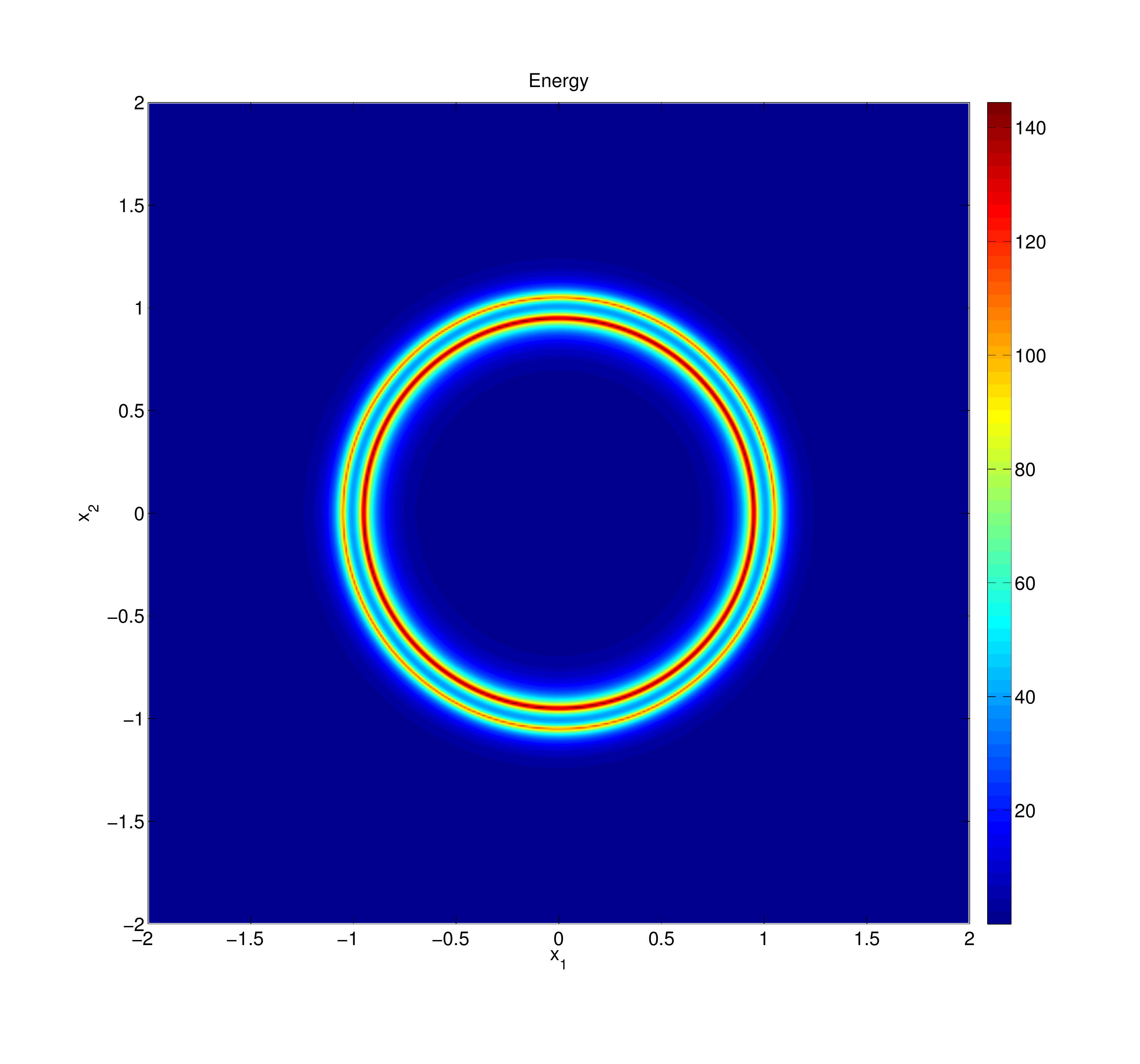}
	}
	\subfigure[] {\label{log-energy-xy}
		\includegraphics[width=60mm,trim=2.3cm 2.5cm 2.5cm 2.8cm, clip=true]{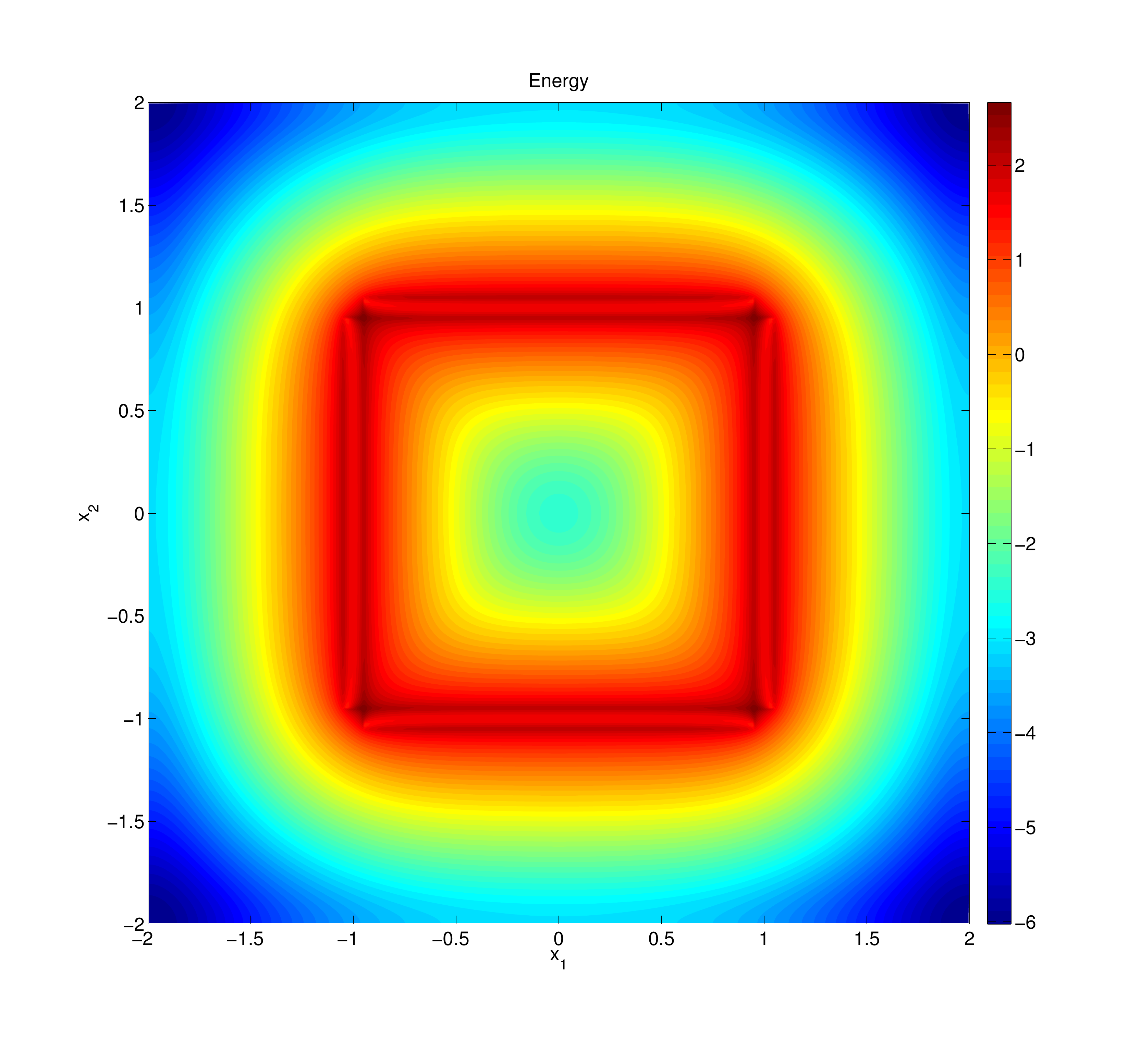}
	}
	\subfigure[] {\label{log-energy-xy-circ}
		\includegraphics[width=60mm,trim=2.3cm 2.5cm 2.5cm 2.8cm, clip=true]{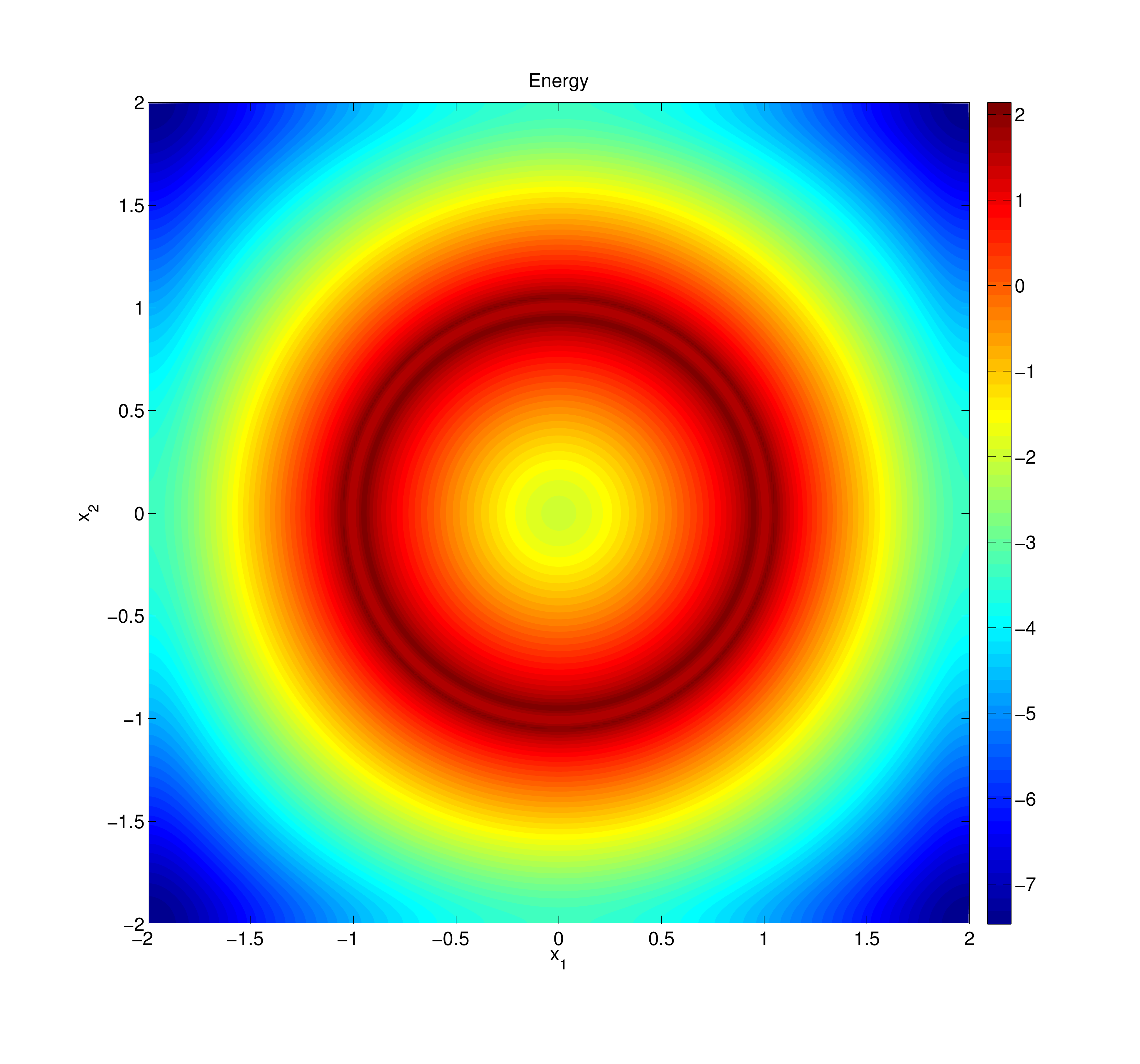}
	}
	\subfigure[] {\label{log-energy-xz}
		\includegraphics[height=24mm, trim=2.2cm 11.9cm 2.7cm 11.9cm, clip=true]{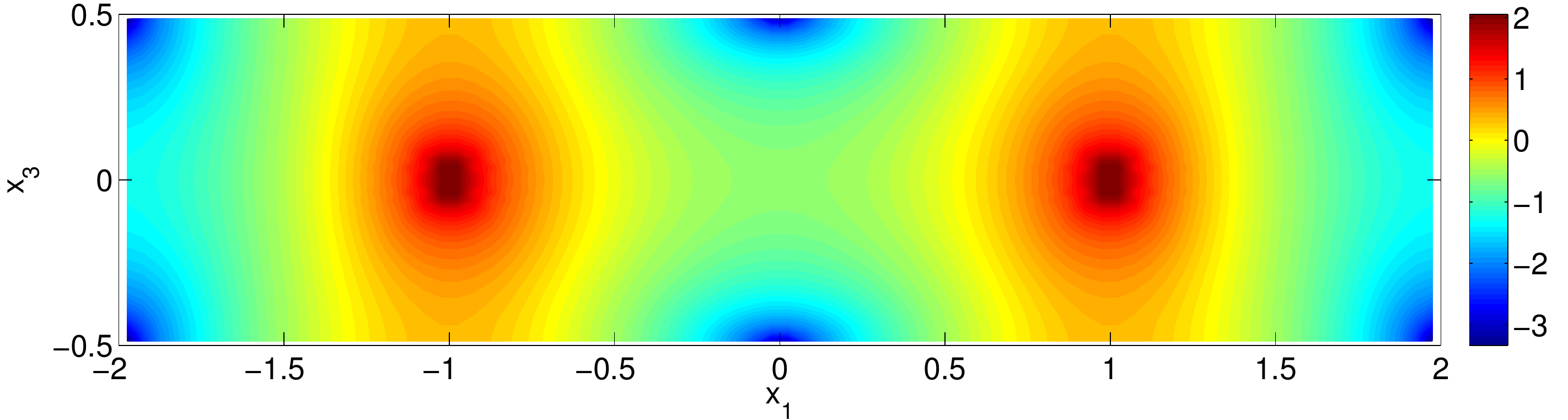}
	}
	\subfigure[] {\label{log-energy-xz-circ}
		\includegraphics[height=25mm, trim=2.2cm 11.9cm 2.7cm 12cm, clip=true]{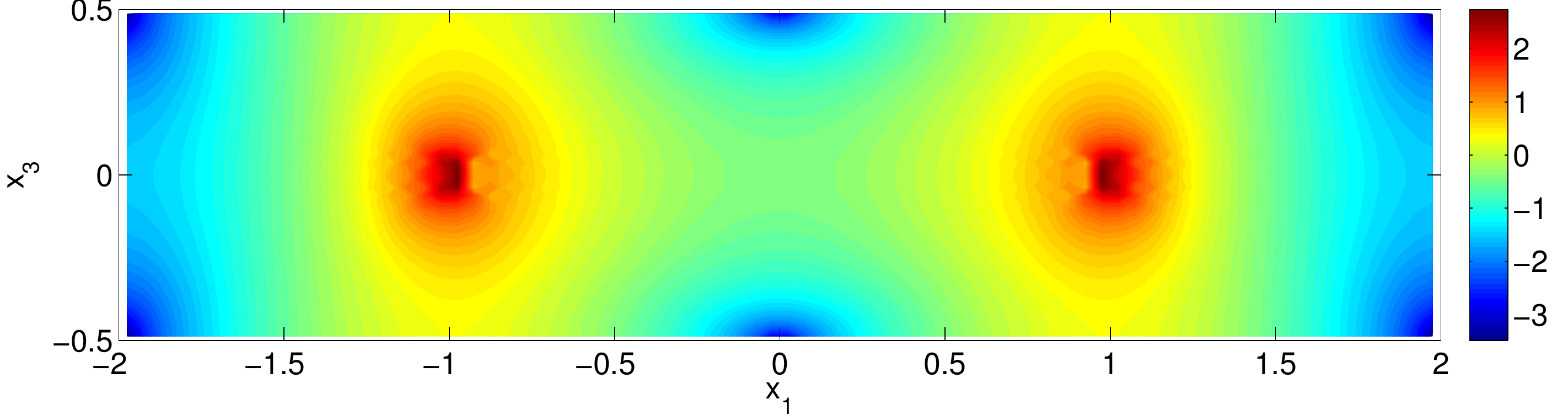}
	}
}
\caption{Energy plot of the twist disclination for square-shaped and circular loops: (a) and (b) are the contours of the normalized energy density, where $x_3=0$, (c) and (d) show the decay of the same contours, in a logarithmic scale. Also, (e) and (f) are the logarithmic contour plots of the energy density in $x_1-x_3$ plane, where $x_2=0$.}
\label{energy-loop}
\end{figure}
\subsection{Convergence of numerical results}
One of the main advantages of our model is the possibility of attaining non-singular results in energy density. In this section we have verified this claim by refining the mesh in our numerical model to demonstrate convergence even inside the defect core. Figure \ref{energy-converge} shows this convergence for the straight wedge disclination. Since the results for different disclination strength are similar, we have normalized our energy plots with respect to $K$.\\
\begin{figure}[!htb]
\centering{
	\subfigure[] {
		\includegraphics[width=50mm,trim=2.cm 2.5cm 2.7cm 2.8cm, clip=true]{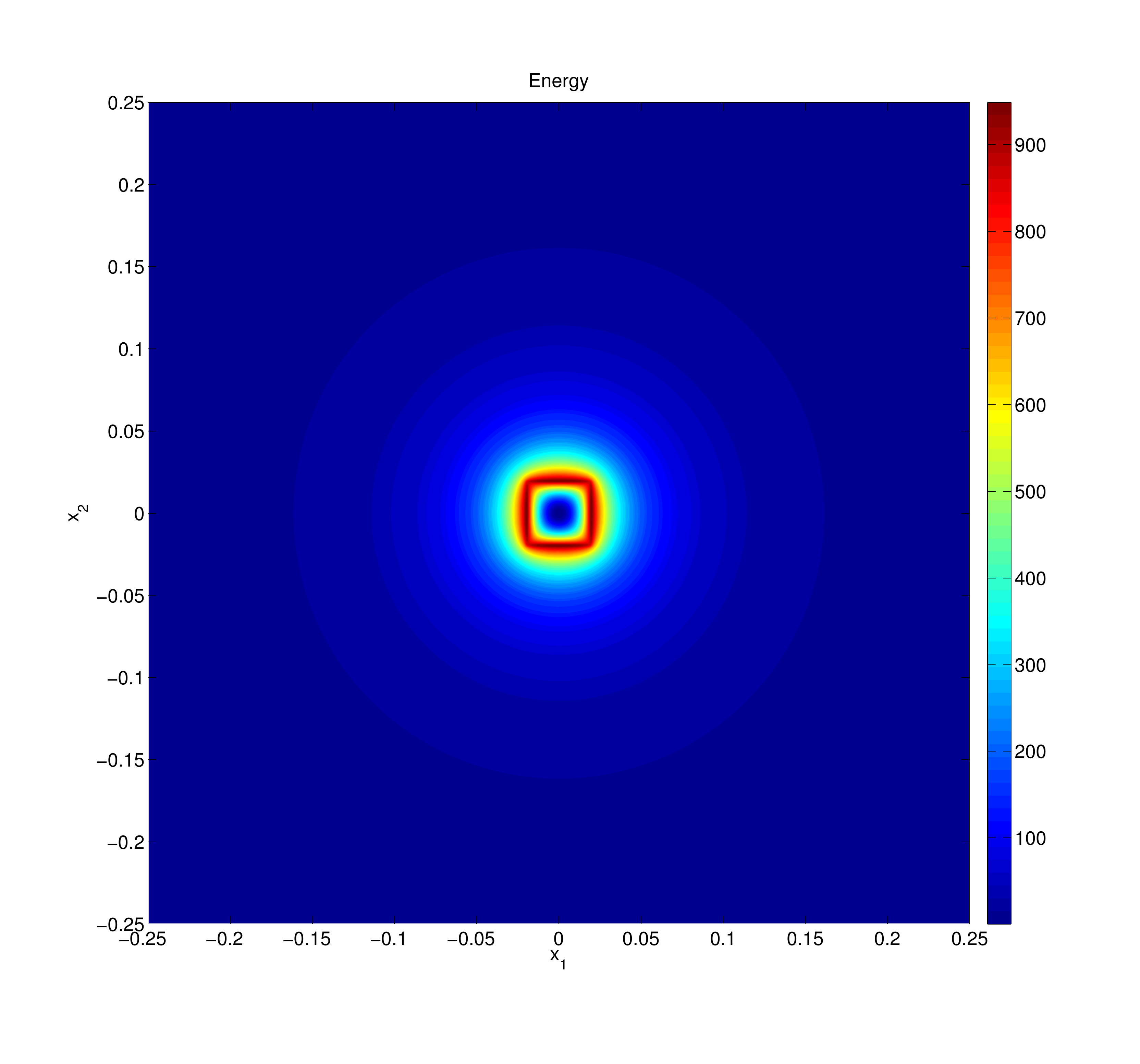}
	}
	\subfigure[] {\label{energy-profile-conv}
		\includegraphics[width=70mm, trim=0.4cm 0.6cm .4cm 0.2cm, clip=true]{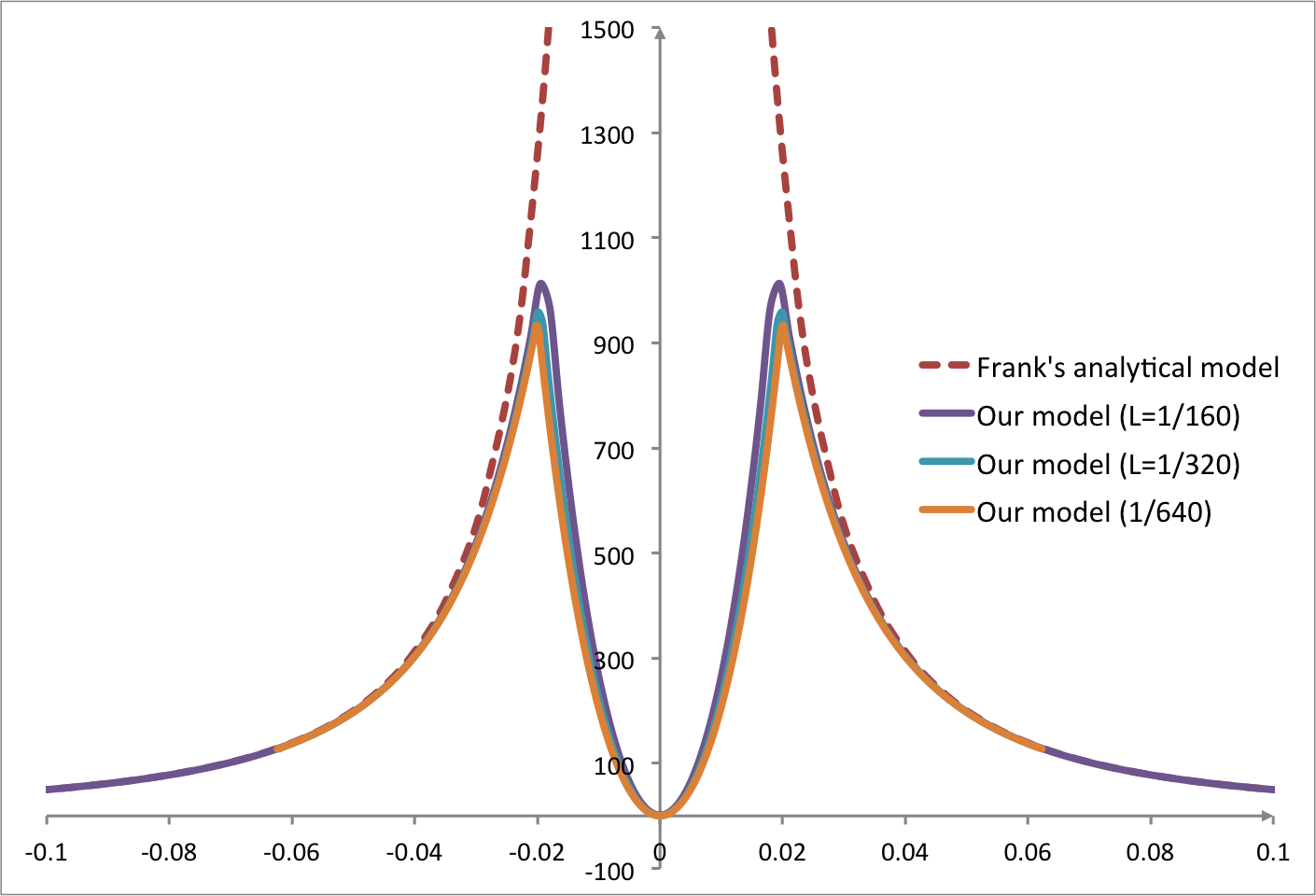}
	}
}
\caption{(a) Normalized energy density contour of wedge disclination including the defect core and (b) The decay of energy density when $x_2=0$, for different element sizes, $L$.}
\label{energy-converge}
\end{figure}
The energy density results for our model and Frank's analytical model are the same, outside the core. Figure \ref{error} shows a relative error of less than $1\%$ everywhere for the energy density, and less than $0.6\%$ for the director field (outside of the layer) for the half-integer defect. Results for all other disclination strengths show the same error distribution. 

\begin{figure}[!htb]
\centering{
	\subfigure[] {
		\includegraphics[width=60mm,trim=1.8cm 2.2cm 1.9cm 1.3cm, clip=true]{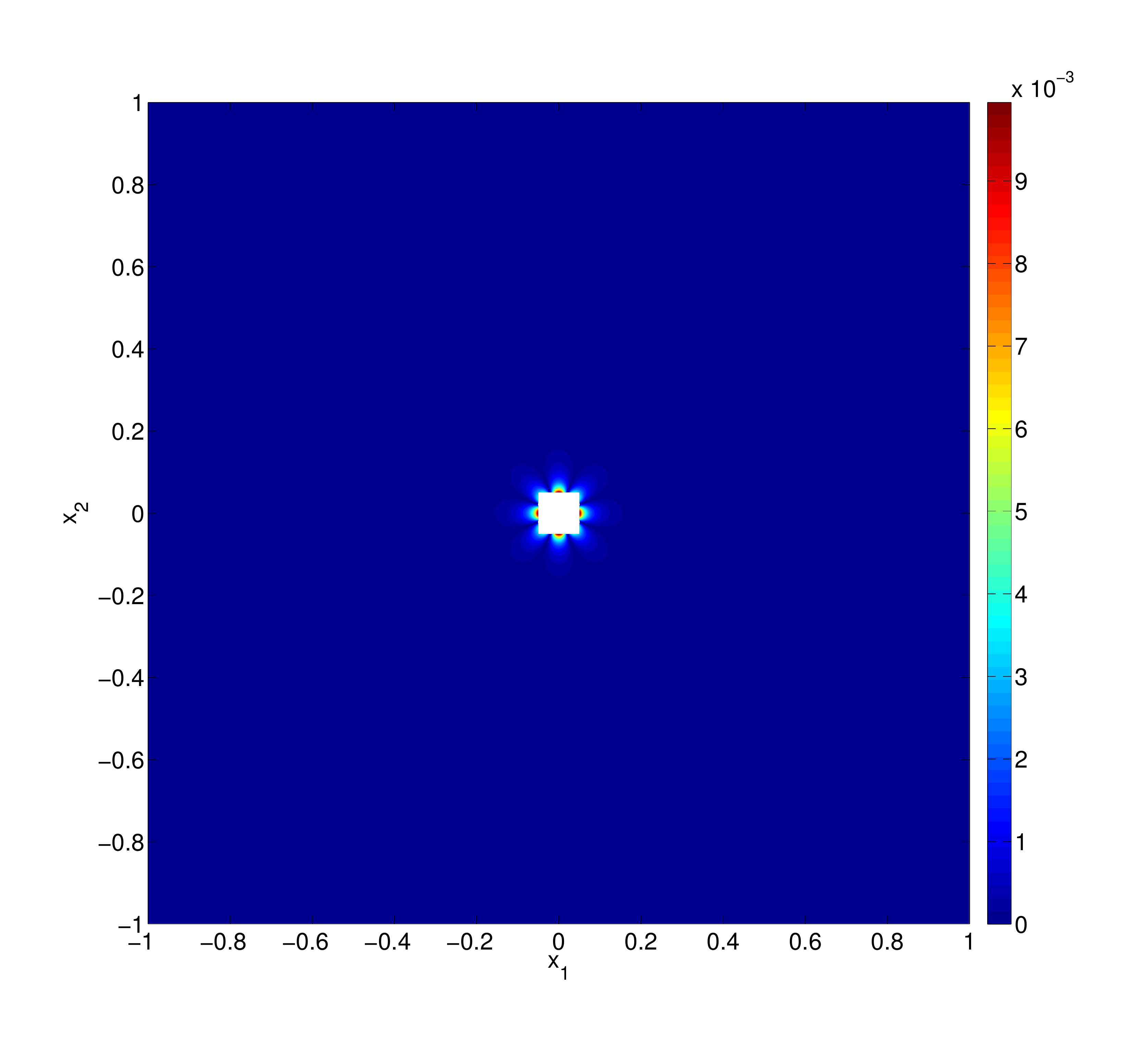}
	}
	\subfigure[] {
		\includegraphics[width=60mm,trim=2cm 2cm 2.2cm 2cm, clip=true]{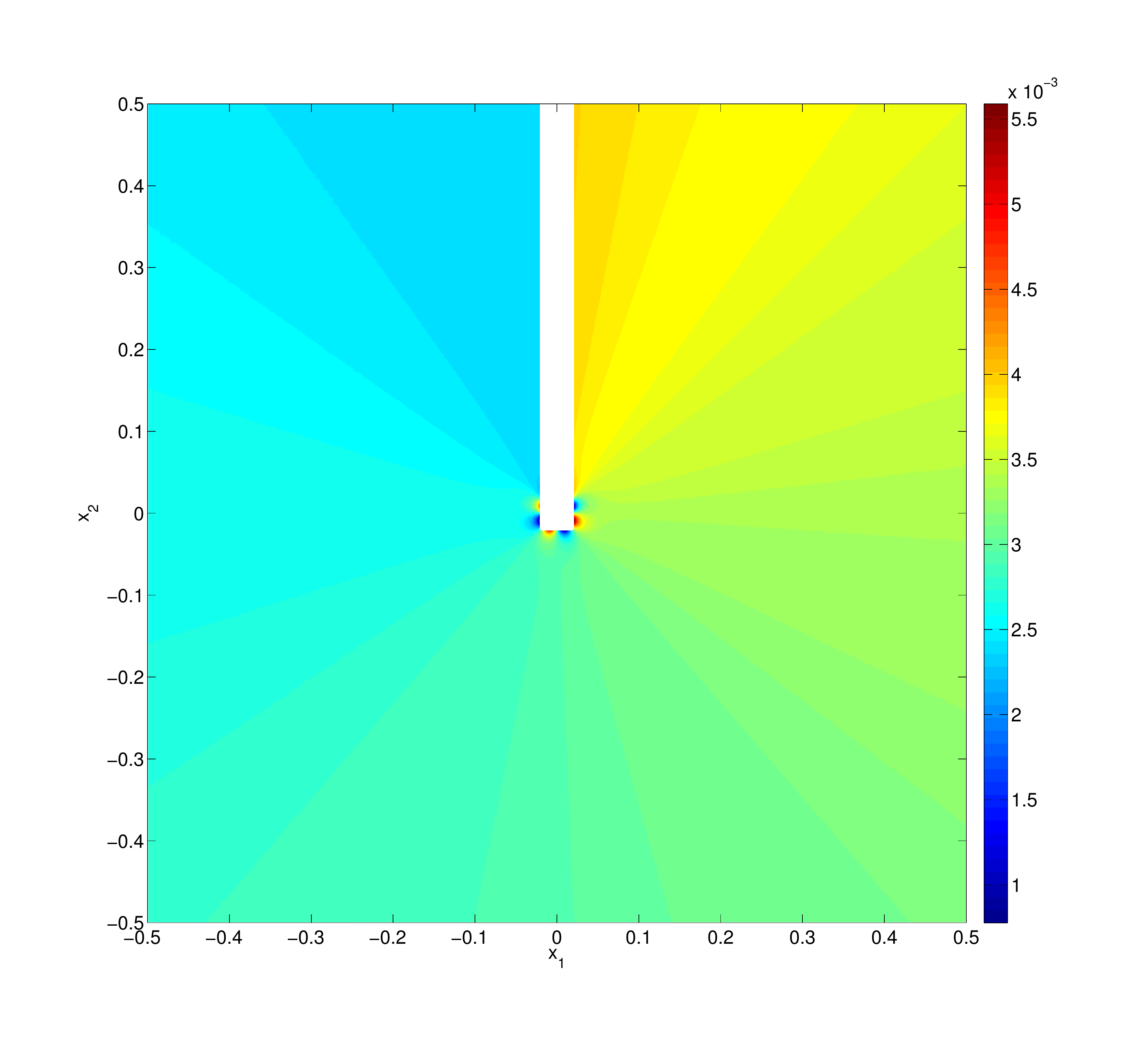}
		\label{error-director}
	}
}
\caption{Comparison of the computed results from the model with Frank's model: (a) Relative error of energy for wedge disclination, (b) Relative error of director field for wedge disclination.}
\label{error}
\end{figure}

Finally, Figure \ref{loop-converge} shows the convergence of the energy density in the square-shaped twist loop model. In this Figure, the energy density profile of the loop with two different element sizes has been compared on the $x_1$ axis, i.e. on the line $x_2=0, x_3=0$.

\begin{figure}[!htb]
\centering{
	\includegraphics[width=70mm,trim=.5cm .1cm .2cm .1cm, clip=true]{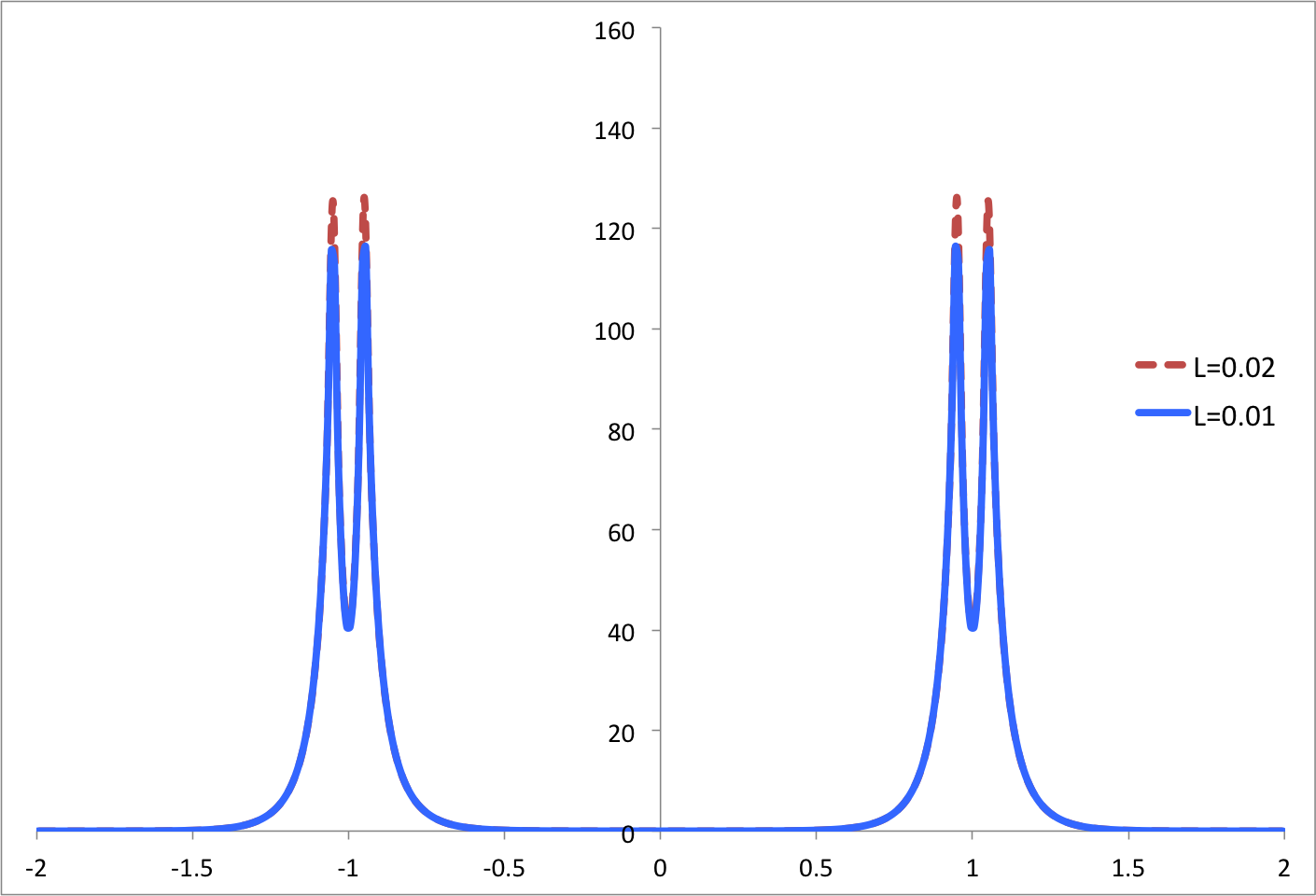}
}
\caption{Normalized energy density profile of the twist disclination for the rectangular loop in Figure \ref{energy-xy}, when $x_2=0$. The element sizes are $L=0.02$ (red, dashed line) and $L=0.01$ (blue, solid line).}
\label{loop-converge}
\end{figure}


\section{Balance of forces}\label{lin-mom}
Assuming a single disclination in the body (loop or a core cylinder terminating at the boundaries of the body), static balance of linear momentum (\ref{force-balance}) is satisfied in the region of the body excluding the core. This is so since the stress tensor is given by
\[
T_{ij} = - p \delta_{ij} - \kappa E^\theta_{i} E^\theta_j
\]
and (\ref{divEtheta}) and (\ref{curlEtheta}) imply, along with the core-support of $\bfbeta^\theta$, that
\[
( E^\theta_{i} E^\theta_j )_{,j} = ( E^\theta_j E^\theta_j )_{,i} = \frac{\psi_{,i}}{\kappa},
\]
and one chooses the indeterminate constraint pressure field in the excluded-core region of the body as
\[
p = - \psi ,
\]
to satisfy (\ref{force-balance}). On the external boundary of the body, recognizing (\ref{bc}) one chooses $p = 0$. Thus, the resultant force on the surface of the core cylinder (or any other surface enveloping it) must vanish.

An interesting possibility is afforded by the model \cite{acharya-dayal} in satisfying static balance of linear momentum everywhere. Note that within the core $curl \bfE^\theta$ does not vanish so that it is not \emph{a priori} clear that $( E^\theta_{i} E^\theta_j )_{,j}$ can be expressed as the gradient of a scalar field. However, the vanishing of the $curl$ of the field $(grad \bfE^\theta) \bfE^\theta$ along with the conditions (\ref{divEtheta}-\ref{beta}-\ref{curlEtheta}) within the core region may be imposed as a design constraint on the $\bfbeta^\theta$ field. Were solutions to exist to this system for $\bfbeta^\theta$, then one could proceed to set up the field $\bflambda$ as before with the caveat that now the field $curl \bflambda$ within the core is not subject only to the mild constraint (\ref{beta}) but instead to the pointwise constraint defined by our calculated $\bfbeta^\theta$ field.

Alternatively, the theory also allows the ansatz
\begin{equation}\label{ansatz2}
\bfE = \sum_{\alpha = 1}^{2} \parderiv{\bfn}{\theta^\alpha} \otimes \left[ grad\, \theta^\alpha - \bflambda^\alpha \right],
\end{equation}
and choosing this affords four more degrees of field degrees of freedom to satisfy static balance of linear momentum without body forces simultaneously with balance of angular momentum - presumably even without necessarily requiring the constraint of incompressibility. Moreover, the solutions generated in this paper would seem to suggest that these extra degrees of freedom could be invoked only in the core cylinder, thus conferring an automatic multiple spatial length scale feature to the theory. Of course, with static balance of linear momentum satisfied everywhere (including cores), the resultant force of the Ericksen stress tensor (and hence the energy-momentum tensor of the classical nematic theory \cite{eshelby1980force, cermelli2002evolution}) on the bounding surface enveloping any segment of a core has to vanish without the requirement of any externally applied body forces or the onset of flow, while the \cite{acharya-dayal} dissipative dynamical theory provides a driving force for disclination motion not related to the Ericksen stress or its resultant. This possibility has the potential of reconciling several different points of view and clarifying an often confusing debate in the liquid crystal literature (cf. \cite{eshelby1980force, kroner1993configurational, ericksen1995remarks}) revolving around the physically meaningful question of modeling the experimentally observed motion of disclination curves in the presence of negligible or no flow.


\section{Concluding Remarks}\label{conclude}
We have demonstrated the possibility of representing non-singular isolated disclination fields within a dynamic theory \cite{acharya-dayal} of liquid crystalline media designed to account for disclination and dislocation defects. The solutions satisfy balance of moments everywhere in the body, including the core. Balance of forces is satisfied outside the cores and it appears possible, by going beyond the ansatz (\ref{ansatz}) assumed in this work, to satisfy force balance everywhere in the body. The fields constructed here may be considered to be equilibria for the theory when the defect lines are considered to be externally-pinned (i.e. the defect velocity field, a constitutive quantity, is assumed to vanish).

Our primary goal has been to demonstrate distinguished pairs of director and (incompatible) director distortion fields $(\bfn,\bfE)$ satisfying (\ref{moment_balance}) and (\ref{curlEtheta}) that generate classically familiar director and energy density fields of disclinations (but not exactly the same ones). In doing so, we use an intermediary `layer' field which, thankfully, is not required in the subsequent dynamics of the \cite{acharya-dayal} model once the director and the director distortion fields have been constructed. However, a fact that gives us pause is that the director incompatibility tensor fields $\bfbeta$ of these isolated disclinations appear to contain some signature of the layer field used to construct them even outside the region that one would typically like to associate with the core of the disclination, and whether this is physically realistic or not remains to be seen. Also, the director fields that we construct have a rapid variation in the layer (without a corresponding energy cost), and the physical realism of such a distribution should also be testable in the modeling of situations involving interaction of the director field with flow. Of course, the analysis presented in Section \ref{argument} provides the logic to construct a director and director incompatibility field without the layer-related features discussed above. To wit, considering the example of the axial wedge disclination of Section \ref{wedge} in an infinite domain, we define
\[
\bflambda_s = grad\, z_s + \bfE^\theta_r
\]
where
\[
div(grad\, z_s) = div\, \bflambda_s
\]
and $\bflambda_s$ is the distribution obtained in (\ref{sing_wall}) by letting $r_1 \rightarrow 0$. This results in $z_s$ being exactly Frank's solution (\ref{frank}). Additionally, we let $\bfE^\theta_r$ be the solution to
\begin{eqnarray}\nonumber
div\, \bfE^\theta_r & = & 0 \\\nonumber
curl\, \bfE^\theta_r  & = & \bfbeta^\theta_r,
\end{eqnarray}
(with natural decay requirements at large distances) where $\bfbeta^\theta_r$ is a radially symmetric core
function described in Appendix \ref{appB} with associated solution for $\bfE^\theta_r$. This formula shows that outside the core, $\bfE^\theta_r$ matches the standard interpretation (\ref{frank-grad}) of the gradient of Frank's solution (\ref{frank}). Setting $\bfE^\theta := grad\,\theta - \bflambda_s$ and requiring it to satisfy the governing equations of our model, i.e. (\ref{divEtheta}) and (\ref{curlEtheta}), a (discontinuous) solution is certainly $\theta = z_s$, with discontinuity supported on the half plane defined by $r_1 \rightarrow 0$ in the definition (\ref{sing_wall}) of $\bflambda_s$. These results imply that outside the core and away from the half-plane $x_1 = 0, x_2 >0$, $\bfE^\theta_r$ matches $\grad\,\theta$. Assuming the validity of (\ref{beta-in-layer}) on the half-plane and noting that the $grad\,\theta$ and $\bfE^\theta$ fields are parallel to $\bfe_1$ on it, it may be concluded that $\bfbeta$ is non-vanishing only inside the core. With $\bfE^\theta$ defined as in this paragraph, we note that even though $grad\,\theta$ is not a locally integrable field, this is exactly compensated by $grad\,z_s$ so that the energy density is formed from the field $\bfE^\theta_r$ alone. In this connection, we note the interesting fact that the classical defect theory works with a director-`gradient' field for forces, moments and energetics that is not strictly a generalized derivative of what it states as the director field (even outside the core)\footnote{The analog of Weingarten's construction to be utilized here makes it entirely clear why this must be so - there, one begins with an irrotational field in a doubly-connected 3-d domain with a through-hole and what may be associated as the `closest' potential field generating the irrotational field is not defined on the same domain but rather on a simply-connected domain obtained through a single cut of the doubly-connected domain that renders the latter simply-connected. Unfortunately, however, these are not procedures that lend themselves to practical nonlinear theory, especially of the type interpretable by the intelligence level of electronic computers that are essential to probe the complex dynamics of the theories we have in mind.}. Thus, it would seem that considering the energy formed from a director distortion field distinct from the director gradient in the presence of line defects (as we do) is probably a reasonable logical device even for discussing the classical theory.

This paper has not made any statements about whether the chosen isolated defect-like director and director distortion fields can actually occur as equilibria of the overall dynamical model \cite{acharya-dayal} for suitable constitutive assumptions for defect velocities (e.g., a linear kinetic assumption on disclination velocity as a function of its theoretical driving force). Based on our experience with related work on Field Dislocation Mechanics \cite{acharya2010}, \cite{acharya-tartar}, \cite{acharya-matthies-zimmer}, we believe this should be possible with the use of non-convexity in OZF energy (possibly due to a limit on director elasticity), the system (as opposed to scalar) nature of the defect evolution equations, as well as from appropriate boundary conditions. In the presence of flow, there are, of course, other natural means to induce line defects.

There are many natural further steps to explore: solving for more-than-planar fields of isolated defects, defect solutions for the full 4-constant Oseen-Frank energy, solutions for disclinations coupled to dislocations in smectics and columnar phases and on to dynamics with and without flow (without flow and even without director inertia, the model has dynamics due to defect evolution). While each of these problems is significant in its own right, we take some satisfaction in the fact that all of these questions can at least be posed unambiguously within our modeling framework. Finally, an important future task is to make contact with the fine topological work of Kleman and Friedel \cite{kleman-friedel} and Kamien and co-workers \cite{Kamien_PNAS}.


\section*{Acknowledgments}
We thank the Department of Civil \& Environmental Engineering at Carnegie Mellon University and the NSF (DMS-0811029; PI: N.J.Walkington) for financial support provided to Hossein Pourmatin. The writing of this paper was accomplished while Pourmatin was supported on NSF grant CAREER-1150002 awarded to K.Dayal.

\section*{Appendices}

\appendix
\section{An Explicit solution for $\theta$}\label{appA}
An analytic solution for the director field corresponding to $\bflambda$ source fields in \ref{Poisson} in an infinite domain is developed, to serve as motivation for some of our claims in Section \ref{argument}. The source fields are layer-like but do not have compact support on bounded domains. Consider  a  $\delta$ sequence \cite{Hancock}:
	\[
	\delta_n(x)=\frac n{\sqrt{\pi}}e^{-(nx)^2}.
	\]
So that, $\int_{-\infty}^{\infty} \delta_n(x)=\erf(x)=1$, for any $n$. Also, consider a $H$ sequence, which converges to the Heaviside step function
	
	\[
	H_n(x)=\frac12\left(\tan^{-1}(nx)+1\right).
	\]
Now, one can define $\bflambda$ as
	\[
	\bflambda_n(\bfx)=2\pi K \delta_n (x_1)H_n(x_2)\bfe_1.
	\]
Note that, for any given $n$,

\[
\oint\bflambda_n.dr=2\pi K\int_{x_1^+}\delta_n(x_1)dx_1=2\pi K
\]
In order to solve (\ref{Poisson}), one can use Green's function method for an infinite 2D domain, where the Green's function is
\[
G(\bfx,\bfx')=\frac1{2\pi}\ln(|\bfx-\bfx'|).
\]
Now, the expression for $\theta_n$ in terms of the Green's function will be
\[
	\theta_n=\int_{V'}G(\bfx,\bfx')div\,\bflambda_n(\bfx')dV
\]
Integrating by part and considering the fact that the $\bflambda_n$ field has only the $\bfe_1$ component
\begin{align} \nonumber
	\theta_n&=\int_{\partial V'}G(\bfx,\bfx')\bflambda_n(\bfx') \cdot \bfnu dS-\int_{V'}\frac{\partial}{\partial x'_1}G(\bfx,\bfx')\bflambda_n(\bfx')dV\\ \nonumber
	&=-\frac1{2\pi}\int_{V'}\frac{x'_1-x_1}{\left|\bfx'-\bfx\right|^2}\bflambda_n(\bfx')dV\\ \nonumber
	&=\frac1{2\pi}\int_{x'_1}\int_{x'_2}\frac{x_1-x'_1}{\left|\bf x'-\bf x\right|^2}2\pi k\delta_n (x'_1)H_n(x'_2)dx'_2dx'_1
	\end{align}
	As a result:
\begin{equation}\label{zn}
	\theta_n=\frac{Kn}{2\sqrt{\pi}}\int_{x'_1}\int_{x'_2}\frac{x_1-x'_1}{\left|\bfx'-\bfx\right|^2}e^{-(nx'_1)^2}\left(\tan^{-1}(nx'_2)+1\right)dx'_2dx'_1
\end{equation}
Note that in case of $n\to\infty$, 

\begin{align} \nonumber
\theta_{\infty}&=\frac{K}2\int_{x'_1}\int_{x'_2}\frac{x_1-x'_1}{\left|\bfx'-\bfx\right|^2}\delta_{\infty}(x'_1)H(x'_2)dx'_2dx'_1 \\ \nonumber
&=\frac K2\int_{x'_2=0}^\infty\frac{x_1}{x_1^2+(x_2-x'_2)^2}dx'_2 \\ \nonumber
&=\frac K2\tan^{-1}\Big(\frac{x_2-x'_2}{x_1}\Big)\Big|_{x'_2=0}^{\infty}
\end{align}
as a result, $\theta_n$ will converge to
\begin{equation}\label{z}
	\theta=K\tan^{-1}\left(\frac{x_2}{x_1}\right)+c,
\end{equation}
which is exactly Frank's solution for planar director fields for straight disclinations for the 1-constant energy.
\section{Solution for $\bfE^\theta_r$}\label{appB}
We provide the following solution from \cite{acharya-dayal} for completeness. Consider $\bfbeta^\theta_r$ given by
\begin{equation}
	\bfbeta^\theta_r (x_1, x_2) = \left\{ \begin{array}{l} \rho(r) \bfe_3, r < r_c  \\ {\bf 0}, r \ge r_c \end{array} ; r = (x_1^2 + x_2^2)^\half \right.
\end{equation}
with the stipulation that
\begin{equation}
 	2 K \pi = \int_0^{2\pi} \int_0^{r_c} \rho(r) \ dr \ r d\psi \Rightarrow K = \int_0^{r_c} \rho(r) \ r dr 
\end{equation}
where $K$ is the strength of the disclination and $r_c$ is a core radius.
Then following Section 5.2 in \cite{Acharya2001}, we have the solution
\begin{equation}
\begin{split}
 	\text{For } r < r_c:& \quad E_{r1}^\theta = \frac{-x_2}{r^2} \int_0^r \rho(s) s \ ds,\ \ \  E_{r2}^\theta = \frac{x_1}{r^2}  \int_0^r \rho(s) s \ ds \\
 	\text{For } r \ge r_c:& \quad E_{r1}^\theta = K \frac{-x_2}{r^2},\ \ \  E_{r2}^\theta = K \frac{x_1}{r^2}. \\
\end{split}
\end{equation}
It can be checked that this solution indeed satisfies the governing equations for $\bfE^\theta_r$ for any $\rho$ satisfying the given conditions.
For the choice
\begin{equation}\nonumber
	\label{eqn:disclination-example}
 	\rho(r) = \left\{ \begin{array}{l} \frac{2K}{r_c} \left(\frac{1}{r} - \frac{1}{r_c}\right), r \le r_c \\ 0, r > r_c \end{array} \right.
\end{equation}
the inside-core solution evaluates to
\begin{equation}\nonumber
 	\text{For } r \le r_c: \quad E_{r1}^\theta = \frac{-x_2}{r^2} \left( \frac{2K}{r_c} \left[ r - \frac{r^2}{2r_c} \right] \right),\ \  E_{r2}^\theta = \frac{x_1}{r^2}  \left( \frac{2K}{r_c} \left[ r - \frac{r^2}{2r_c} \right] \right) \\
\end{equation}
while the outside-core distribution remains unchanged.

\bibliographystyle{alpha} \bibliography{disclinations}

\end{document}

%% file: temp-definitions.tex
\usepackage{amsmath}
\usepackage{amsbsy}
\usepackage{amssymb}
\usepackage{amscd}
\usepackage{amsfonts}

\newcommand{\bfe}{{\mathbold e}}

\newcommand{\bfg}{{\mathbold g}}

\newcommand{\bfn}{{\mathbold n}}

\newcommand{\bfx}{{\mathbold x}}

\newcommand{\bfA}{{\mathbold A}}
\newcommand{\bfB}{{\mathbold B}}

\newcommand{\bfE}{{\mathbold E}}

\newcommand{\bfT}{{\mathbold T}}

\newcommand{\beq}{\begin{equation}}
\newcommand{\eeq}{\end{equation}}
\newcommand{\beqs}{\begin{eqnarray}}
\newcommand{\eeqs}{\end{eqnarray}}
\newcommand{\beql}{\begin{equation} \label}
\newcommand{\half}{\frac{1}{2}}

\newcommand{\bfnu}{\mathbold{\nu}}

\newcommand{\bfbeta}{\mathbold{\beta}}
\newcommand{\bflambda}{\mathbold{\lambda}}

\newcommand{\erf}{\mathop{\rm erf}\nolimits}

\newcommand{\grad}{\mathop{\rm grad}\nolimits}

\newcommand{\curl}{\mathop{\rm curl}\nolimits}

\newcommand{\parderiv}[2]{\frac{\partial #1}{\partial #2}}